\documentclass[a4paper]{article}
\usepackage{setspace}
\usepackage[margin=1in]{geometry}
\usepackage{amsmath,amsfonts,amssymb,graphicx,enumitem, booktabs, tikz, xcolor}
\usepackage[ruled,linesnumbered]{algorithm2e}
\usepackage[sectionbib]{natbib}
\usepackage[flushleft]{threeparttable}
\usepackage{rotating}
\usepackage{amsthm}
\usepackage{float}
\usepackage{placeins}

\allowdisplaybreaks[2]
\definecolor{mypurple}{RGB}{160, 32, 240}
\providecommand{\keywords}[1]
{
  \small	
  \textbf{\textit{Keywords:}} #1
}
\newtheorem{theorem}{Theorem}
\newtheorem{condition}{Condition}
\newtheorem{proposition}{Proposition}
\theoremstyle{remark}
\newtheorem{remark}{Remark}

\newsavebox{\tablecontent}
\newcommand{\tbl}[2]{\caption{#1}\centering
  \sbox{\tablecontent}{#2}%
  \ifdim\wd\tablecontent>\linewidth
    \resizebox{\linewidth}{!}{\usebox{\tablecontent}}%
  \else\usebox{\tablecontent}\fi\par}
\newenvironment{tabnote}{\par\smallskip\begin{minipage}{\linewidth}\footnotesize}{\end{minipage}\par}
\allowdisplaybreaks[2]

\begin{document}

\title{An optimal two-step estimation approach for two-phase studies}

\author{
Qingning Zhou\textsuperscript{1} and Kin Yau Wong\textsuperscript{2,3}\thanks{Corresponding author; email: kin-yau.wong@polyu.edu.hk} \\
\textsuperscript{1}Department of Mathematics and Statistics,\\
The University of North Carolina at Charlotte\\
\textsuperscript{2}Department of Applied Mathematics, The Hong Kong Polytechnic University\\
\textsuperscript{3}Hong Kong Polytechnic University Shenzhen Research Institute
}

\date{}

\maketitle

\begin{abstract}
Two-phase sampling is commonly adopted for reducing cost and improving estimation efficiency. In many two-phase studies, the outcome and some cheap covariates are observed for a large sample in Phase I, and expensive covariates are obtained for a selected subset of the sample in Phase II. As a result, the analysis of the association between the outcome and covariates faces a missing data problem. Complete-case analysis, which relies solely on the Phase II sample, is generally inefficient. In this paper, we study a two-step estimation approach, which first obtains an estimator using the complete data, and then updates it using an asymptotically mean-zero estimator obtained from a working model between the outcome and cheap covariates using the full data. This class of two-step estimators are asymptotically at least as efficient as the complete-data estimator and are robust to misspecification of the working model. We propose a kernel-based method to construct a two-step estimator that achieves optimal efficiency within this class of estimators. Additionally, we develop a simple joint update approach based on multiple working models to approximate the optimal estimator when a fully nonparametric kernel approach is infeasible. 
We illustrate the proposed methods with various outcome models. We demonstrate their advantages over existing approaches through simulation studies and provide an application to a major cancer genomics study.
\end{abstract}


\section{Introduction}

Two-phase sampling is a widely used technique that aims at cost reduction and improved efficiency of estimation. Typically, in Phase I, a large sample is drawn from a target population, and variables that are cheap or easy to obtain are measured. These variables could include outcomes, cheap covariates, and auxiliary variables that are correlated with more expensive covariates not available in Phase I. They  can be used to define strata within the Phase I sample. In Phase II, a subsample is drawn from each stratum to measure variables that are expensive or difficult to obtain, such as biomarkers ascertained by bioassay or genetic analysis and medical records that rely on labor-intensive chart review. The introduction of strata seeks either to oversample subjects with important Phase I variables, or to effectively sample subjects with targeted Phase II variables, or both. Thus, two-phase sampling achieves efficient access to important variables with less cost. For example, the case-cohort design, initially proposed by \citet{prentice1986case}, is a widely used two-phase sampling design for cost reduction in large cohort studies that concern rare diseases and expensive covariates. Under this design, the measurements of expensive covariates are ascertained only for a random sample of the cohort, referred to as the subcohort, and for all subjects who were observed to have developed the disease during the study period, referred to as cases. There is an extensive literature on analyzing data from the case-cohort design and its variations, such as stratified or generalized case-cohort designs \citep{chen1999case,kulich2004improving,cai2007power,kong2009case,kim2013more,zhou2017case}.

The two-phase design has been adopted for some major biomedical studies, such as the Atherosclerosis Risk in Communities (ARIC) study \citep{aric1989atherosclerosis}. The ARIC study is an ongoing longitudinal epidemiological study conducted at four field centers in the United States, where over 15,000 subjects were recruited in 1987 and have been periodically examined thereafter. The subjects were followed for some disease outcomes of interest, such as coronary heart disease, stroke, diabetes, hypertension, and death. Some expensive covariates, such as high-sensitivity C-reactive protein, DNA alterations, DNA methylation, and metabolites, were only collected for a case-cohort sample or a random sample of the full cohort. The two-phase design has also been adopted in major HIV/AIDS studies. For example, the two Antibody Mediated Prevention trials conducted by HIV Vaccine Trials Network (HVTN), HVTN 703 in sub-Saharan Africa and HVTN 704 in Americas and Europe, were designed to investigate whether the broadly neutralizing antibody (bnAb), VRC01, could prevent HIV-1 acquisition \citep{seaton2023pharmacokinetic,corey2021two}. In these trials, the VRC01 measurements were collected only for a case-cohort sample selected from the full cohort. Similar missing-data patterns could also arise from multi-omics studies, such as The Cancer Genome Atlas (TCGA) \citep{cancer2011integrated}. In this study, cancer patients were measured on multiple omics platforms, including DNA alterations and the expressions of RNA and protein, at different locations and time points. While most of the subjects were measured on DNA and RNA, protein expression data were only available for a subset of subjects because limited tissue samples remained when the protein expression platform was introduced.

The two-phase design creates a missing data problem for the analysis of association between the outcome and covariates. A naive method to handle missing data is the complete-case analysis, where subjects with missing observations are discarded. For two-phase studies, the complete-case analysis considers only subjects selected in Phase II. Under missing completely at random (MCAR), the complete-case analysis using standard approaches yields valid estimation, while under missing at random (MAR), some adjustments, such as inverse-probability weighting (IPW), are required to yield consistent estimation. In any case, the complete-case analysis is generally inefficient, as it fails to utilize information from incomplete observations.

There are several general approaches to deal with missing covariates that make use of information beyond the complete observations. One such approach is maximum likelihood estimation (MLE), where the likelihood incorporates the outcome model and the model of the incomplete covariates \citep{zeng2014efficient,tao2017efficient}. Although the MLE is efficient, a major drawback is that estimation may be inconsistent if the incomplete covariate model is misspecified. Model assumptions can be relaxed by adopting a fully nonparametric model for the incomplete covariates, but this is feasible only when the number of covariates is small. Another approach is (multiple) imputation, where the missing covariate values are imputed based on the observed data and then conventional analyses can be performed on the imputed data \citep{little2019statistical}. Like MLE, misspecified imputation models could lead to inconsistent estimation. Also, estimators derived from imputed data often lack theoretical guarantees. In fact, imputation methods are often viewed as black-box approaches with estimation and inference conducted independently of the imputation step. Simulation studies by \citet{cannings2022correlation} show that multiple imputation can sometimes perform worse than the complete-case analysis. 

Another approach, which is the primary focus of this paper, is the two-step ``update estimation'' approach. This approach involves first constructing an original estimator based on the complete observations and then updating it using a zero-consistent estimator obtained from a working model between the outcome and cheap covariates or auxiliary variables using all study subjects. An advantage of this approach is that the update estimator has (asymptotic) efficiency higher than or equal to that of the original estimator, regardless of whether the working model is correctly specified or not. The update estimation approach, in its current general form for regression models, was first formulated by \citet{chen2000unified} under MCAR. This approach has thereafter been extensively studied under various settings, such as for semiparametric regression models \citep{wang2015semiparametric,wang2018semiparametric} and for censored failure times under different models and data structures \citep{chen2002cox,jiang2007additive,li2008non,li2010semiparametric,liu2010cox,wang2015semiparametric,yan2017improving,zhou2024improving}. Recently, \citet{cannings2022correlation} studied the update estimation approach under more general missing-data patterns for both parametric and nonparametric estimation problems. Similar update approaches have also been developed for incorporating auxiliary information from historical data, external databases, and electronic health records \citep{lin2014adjustment,tong2020augmented,davidov2024use}.

An approach closely related to the update approach is augmented inverse-probability weighting (AIPW), which adds an augmentation term to the inverse-probability weighted estimating equations to improve robustness and efficiency upon the IPW estimator \citep{robins1994estimation}. The AIPW approach possesses the ``double robustness'' property, where the estimator is consistent if either the missing mechanism or the working model underlying the augmentation term is correctly specified. Also, the AIPW estimator is more efficient than the IPW estimator if the augmentation term is correctly specified, but it could be less efficient than IPW for a poorly chosen augmentation term. Some versions of AIPW modify the standard augmentation term to guarantee efficiency gain over IPW \citep{han2012note}, resulting in an estimator asymptotically equivalent to the update estimator. Nevertheless, AIPW may require solving a complex system of equations that needs to be developed for each specific problem.

Although the update estimation approach has been studied and adopted in various settings, there is little work on the implementation or even formulation of the optimal choice of the working model. Typically, the working model is conveniently chosen to be of the same class as the outcome model of the original estimator. While the update estimator is guaranteed to yield efficiency gain over the original estimator regardless of the specification of the working model, it is highly desirable to choose a working model that yields (approximate) optimal efficiency gain. In this paper, we develop novel estimation approaches to achieve optimal efficiency within the class of update estimators in two-phase studies. We derive a general form of the optimal update term and propose a nonparametric estimation method to achieve optimal efficiency. We also develop a simple joint update approach that uses multiple working models to perform the update, which helps approximate the optimal update when the nonparametric estimation is infeasible. The proposed methods are based on the influence function and are generally applicable as long as the original estimator is asymptotically linear. We illustrate the proposed methods using various outcome models, including the linear regression model, logistic regression model, and the Cox proportional hazards model for right-censored survival times. %

\section{Methods}\label{sec2}
\subsection{Preliminaries}\label{subsec1}

\begin{figure}[htbp]
	\begin{center}
		\includegraphics[scale=0.55]{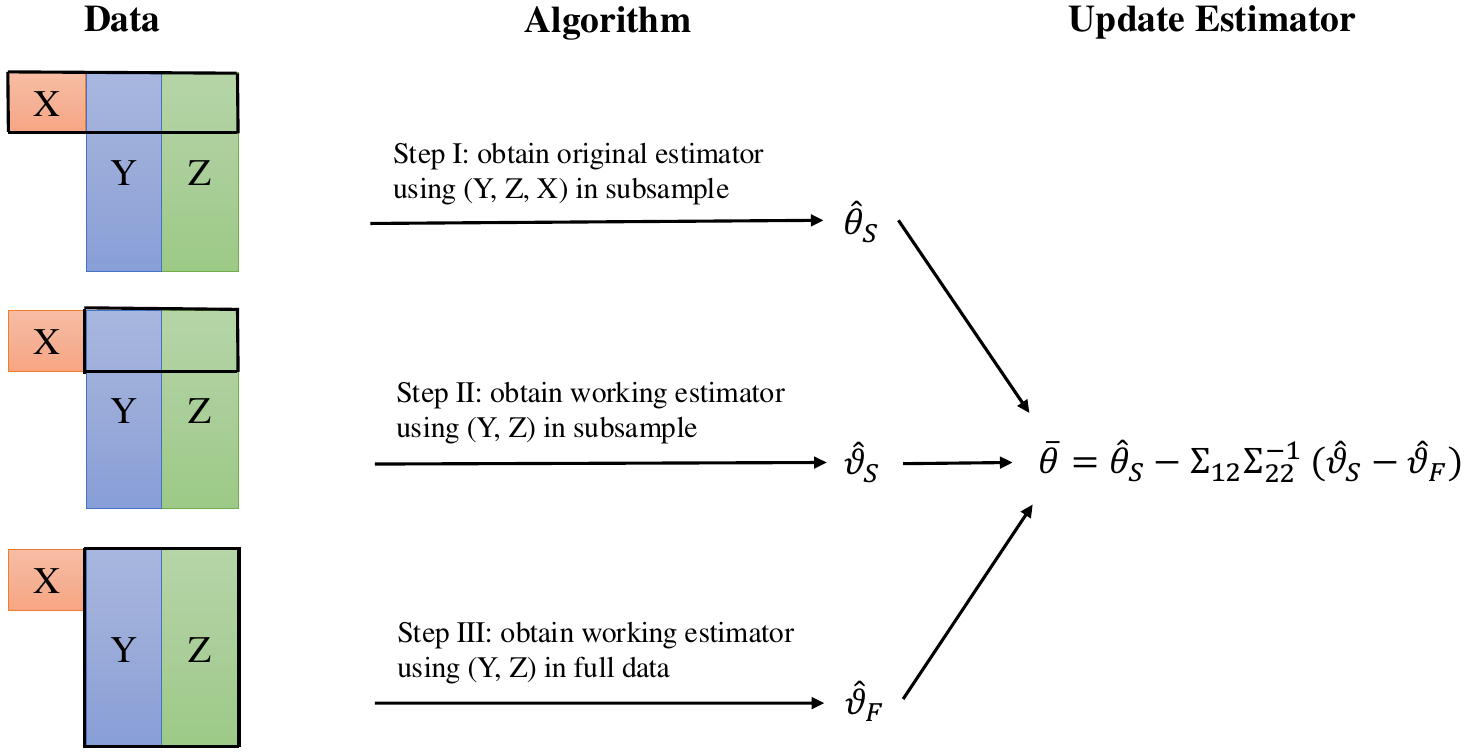}
		\caption[]{Overview of the update estimation approach}
		\label{fig_workflow}
	\end{center}
\end{figure}

We first describe the general form of the update estimation approach for regression analysis with missing covariates. Let $Y$ denote the outcome, $X$ denote a vector of expensive covariates subject to missingness, and $Z$ denote a vector of cheap covariates that are completely observed. We allow $Z$ to include auxiliary variables that are not of direct interest but can act as surrogates for $X$. For survival data, $Y$ may consist of both the observed time and event indicator. Let $R$ be the indicator of whether $X$ is observed. For a sample of size $n$, the observed data consist of $(Y_i,Z_i,R_i,R_iX_i)$ for $i=1,\ldots,n$. Under a two-phase design, $R$ indicates whether a subject is selected into the Phase II sample, and it generally depends on the Phase I data. Specifically, we assume MAR, such that $R$ is independent of $X$ given $(Y,Z)$. In the sequel, we use the term {\it subsample} to refer to the subjects with $R=1$ (i.e., the Phase II sample). %
Let $\theta$ denote a vector of parameters of interest in the model of $Y$ given $(X,Z)$ and $\eta$ denote a set of possibly nonparametric nuisance parameters. Note that this model may not include all components of $Z$, some of which may be auxiliary variables for $X$. Let $\widehat{\theta}_S$ be a consistent estimator of $\theta$ based on the subsample, which can be the MLE under MCAR or an IPW estimator under MAR. Let $\widehat{\vartheta}_S$ be an estimator computed from $(Y_i,Z_i)$ for $R_i=1$ and $\widehat{\vartheta}_F$ be an estimator computed from $(Y_i,Z_i)$ for $i=1,\ldots,n$ that converges to the same limit as $\widehat{\vartheta}_S$. In the literature, $\widehat{\vartheta}_S$ and $\widehat{\vartheta}_F$ are typically chosen as estimators of some parameter $\vartheta$ in a working model of $Y$ on $Z$, based on the subsample and the full sample, respectively. The update estimator is defined as
\begin{align}
\overline{\theta}=\widehat{\theta}_S-\Sigma_{12}\Sigma_{22}^{-1}(\widehat{\vartheta}_S-\widehat{\vartheta}_F), 
\label{eq:update}
\end{align}
where $\Sigma_{12}$ is the asymptotic covariance between $n^{1/2}(\widehat{\theta}_S-\theta_0)$ and $n^{1/2}(\widehat{\vartheta}_S-\widehat{\vartheta}_F)$, $\Sigma_{22}$ is the asymptotic variance of $n^{1/2}(\widehat{\vartheta}_S-\widehat{\vartheta}_F)$, and $\theta_0$ is the true value of $\theta$. We can replace $\Sigma_{12}$ and $\Sigma_{22}$ by their consistent estimators. The data structure and overview of the update estimation approach are depicted in Fig.~\ref{fig_workflow}. Obviously, $\overline{\theta}$ is a consistent estimator of $\theta$. The update estimator utilizes the correlation between $\widehat{\theta}_S$ and the zero-consistent statistic $(\widehat{\vartheta}_S-\widehat{\vartheta}_F)$ to improve efficiency. The asymptotic variance of $\overline{\theta}$ is guaranteed to be smaller than or equal to that of $\widehat{\theta}_S$.

\subsection{Optimal Update}

We study the choice of $(\widehat{\vartheta}_S,\widehat{\vartheta}_F)$ that yields optimal efficiency for the update estimator.
We assume that $\widehat{\theta}_S$ is asymptotically linear and takes the form
\begin{align*}
n^{1/2}(\widehat{\theta}_S-\theta_0) =  n^{-1/2} \sum_{i=1}^n \frac{R_i}{\pi_i} \, \psi(Y_i,X_i,Z_i;\theta_0,\eta_0) + o_p(1),
\end{align*}
where $\pi_i=P(R_i=1\mid Y_i,Z_i)$, $\eta_0$ is the true value of $\eta$, and $\psi(Y,X,Z;\theta,\eta)$ is such that its expectation is zero at $(\theta_0,\eta_0)$. Here, $\widehat{\theta}_S$ is computed using the subsample, with inverse-probability weights to account for potential sampling bias. In general, the distributions of the working estimators $\widehat{\vartheta}_S$ and $\widehat{\vartheta}_F$ depend on $\theta$, $\eta$, and the  conditional distribution function of $X$ given $Z$, denoted by $F$, such that $F(x\mid Z)=P(X\le x\mid Z)$.
We assume that $\widehat{\vartheta}_S$ and $\widehat{\vartheta}_F$ are asymptotically linear with
\begin{align*}
n^{1/2}(\widehat{\vartheta}_S-\vartheta_0) =&\, n^{-1/2} \sum_{i=1}^n \frac{R_i}{\pi_i} \, \phi(Y_i,Z_i;\theta_0,\eta_0,F_0) + o_p(1)
\end{align*}
and
\begin{align*}
n^{1/2}(\widehat{\vartheta}_F-\vartheta_0) = n^{-1/2} \sum_{i=1}^n\phi(Y_i,Z_i;\theta_0,\eta_0,F_0)+ o_p(1), 
\end{align*}
where $\vartheta_0$ is the limit of $\widehat{\vartheta}_S$ and $\widehat{\vartheta}_F$, $(\eta_0,F_0)$ are the true values of $(\eta,F)$, and $\phi(Y,Z;\theta,\eta,F)$ is such that its expectation is zero at $(\theta_0,\eta_0,F_0)$. Here, $\widehat{\vartheta}_S$ is weighted similarly as $\widehat{\theta}_S$ to account for sampling bias. %

Let $\psi_0=\psi(Y,X,Z;\theta_0,\eta_0)$ and $\phi_0=\phi(Y,Z;\theta_0,\eta_0,F_0)$. We assume the following two conditions.

\begin{condition}
The variables $\psi_0$ and $\phi_0$ are mean 0, $\mathrm{E}(\psi_0\psi_0^{\mathrm{T}})<\infty$, and both $\mathrm{E}(\phi_0\phi_0^{\mathrm{T}})$ and $\mathrm{E}\{\pi^{-1}(1-\pi)\phi_0\phi_0^{\mathrm{T}}\}$ are finite and invertible. Also, the $o_p(1)$ terms in the linear expansions of $\widehat{\theta}_S$, $\widehat{\vartheta}_S$, and $\widehat{\vartheta}_F$ converge to 0 in $L_2(\mathbb{P})$.
\end{condition}
\begin{condition}
The probability $P(R=1\mid Y,Z)>\delta$ for some positive constant $\delta$ almost surely.
\end{condition}

Let
\[
\phi^*(Y,Z;\theta,\eta,F) = \frac{\int\psi(Y,x,Z;\theta,\eta)f(Y\mid x,Z;\theta,\eta)\,\mathrm{d}F(x\mid Z)}{\int f(Y\mid x,Z;\theta,\eta)\,\mathrm{d}F(x\mid Z)},
\]
where $f$ denotes the conditional density or probability mass function of $Y$ given $(X,Z)$ evaluated at $(\theta,\eta)$. Note that $\phi^*(Y,Z;\theta,\eta,F)$ is the conditional expectation of $\psi(Y,X,Z;\theta,\eta)$ given $(Y,Z)$, with the parameters in the conditional expectation evaluated at $(\theta,\eta,F)$. Let $\phi^*_0=\phi^*(Y,Z;\theta_0,\eta_0,F_0)$. We have the following results about the efficiency of update estimators.

\begin{theorem}
Under Conditions 1 and 2, we have 
$$
n^{1/2}(\overline{\theta}-\theta_0)\rightarrow_d\mathrm{N}(0,\Sigma(\phi_0)), 
$$
where
\[
\Sigma(\phi_0)=\mathrm{E}\bigg(\frac{1}{\pi}\psi_0\psi_0^{\mathrm{T}}\bigg)-\mathrm{E}\bigg(\frac{1-\pi}{\pi}\psi_0\phi_0^{\mathrm{T}}\bigg)\mathrm{E}\bigg(\frac{1-\pi}{\pi}\phi_0\phi_0^{\mathrm{T}}\bigg)^{-1} \mathrm{E}\bigg(\frac{1-\pi}{\pi}\phi_0\psi_0^{\mathrm{T}}\bigg),
\]
with $\pi=P(R=1\mid Y,Z)$. Also, assuming that $\mathrm{E}(\phi^*_0\phi^{*\mathrm{T}}_0)$ is finite and invertible, we have
\[
\Sigma(\phi^*_0)\preceq\Sigma(\phi_0)
\]
for any $\phi_0$ satisfying Condition 1, where $A\preceq B$ denotes that $B-A$ is positive semidefinite.
\end{theorem}
The proof of Theorem 1, along with those of the subsequent theorems, are given in the Appendix.

From Theorem 1, we see that the optimal choice of $\phi$ is the projection of $\psi$ onto the space of functions of $(Y,Z)$, up to a deterministic invertible linear transformation. Note that $\phi_0$ is defined as an influence function, but with a slight abuse of notation, $\phi^*_0$ is generally not an influence function before applying the linear transformation. For a random sample of size $n$ in general, to yield an estimator of $\theta$ with a given influence function $\phi(\mathcal{X};\theta)$, where $\mathcal{X}$ denotes the observed data for a generic subject, we can set the estimator to be the solution to $\sum_{i=1}^n\phi(\mathcal{X}_i;\theta)=0$. Therefore, if there were no nuisance parameters, then we can set $\widehat{\vartheta}_S$ and $\widehat{\vartheta}_F$ to be Z-estimators that solve empirical (weighted) means of $\phi^*$ equal 0. In general, however, $\eta$ and $F$ are unknown. Let $\widehat{\eta}$ and $\widehat{F}$ denote their respective estimators based on the subsample. We propose to set $\widehat{\vartheta}_S$ and $\widehat{\vartheta}_F$ as the solutions for $\theta$ to the equations
\begin{align}
\sum_{i=1}^n \frac{R_i}{\pi_i} \frac{\int\psi(Y_i,x,Z_i;\theta,\widehat{\eta})f(Y_i\mid x,Z_i;\theta,\widehat{\eta})\,\mathrm{d}\widehat{F}(x\mid Z_i)}{\int f(Y_i\mid x,Z_i;\theta,\widehat{\eta})\,\mathrm{d}\widehat{F}(x\mid Z_i)}=0 
\label{eq:optimal2}
\end{align}
and
\begin{align}
\sum_{i=1}^n \frac{\int\psi(Y_i,x,Z_i;\theta,\widehat{\eta})f(Y_i\mid x,Z_i;\theta,\widehat{\eta})\,\mathrm{d}\widehat{F}(x\mid Z_i)}{\int f(Y_i\mid x,Z_i;\theta,\widehat{\eta})\,\mathrm{d}\widehat{F}(x\mid Z_i)}=0, 
\label{eq:optimal3}
\end{align}
respectively.

We consider the following conditions.
\begin{condition}
The parameter space for $\theta$, denoted by $\Theta$, is compact, and $\theta_0$ is in the interior of $\Theta$.
\end{condition}
\begin{condition}
The function $\phi^*$ is three-times differentiable with respect to $\theta$, with uniformly bounded derivatives over $(Y,Z)$ and $(\eta,F)$. Let the first and second derivatives be $\phi^*_\theta$ and $\phi^*_{\theta\theta}$, respectively. We have
\begin{align*}
\sup_{\theta\in\Theta}\Big|n^{1/2}(\mathbb{P}_n-\mathbb{P})g(Y,Z;\theta,\eta_0,F_0)\Big|=&\,O_p(1)\\
\sup_{\theta\in\Theta}\Big|n^{1/2}(\mathbb{P}_n-\mathbb{P})\Big\{g(Y,Z;\theta,\widehat{\eta},\widehat{F})-g(Y,Z;\theta,\eta_0,F_0)\Big\}\Big|=&\,o_p(1)\\
\sup_{\theta\in\Theta}\Big|\mathbb{P}\Big\{g(Y,Z;\theta,\widehat{\eta},\widehat{F})-g(Y,Z;\theta,\eta_0,F_0)\Big\}\Big|=&\,O_p(r_n)
\end{align*}
for $g=\phi^*$, $\phi^*_\theta$, and $\phi^*_{\theta\theta}$, where $\mathbb{P}$ and $\mathbb{P}_n$ denote the true measure and empirical measure, respectively, and $r_n$ is a positive sequence such that $n^{1/2}r_n^3\rightarrow0$.
\end{condition}
\begin{condition}
The value $\theta_0$ is the unique solution to
\[
\mathbb{P}\phi^*(Y,Z;\theta,\eta_0,F_0)=0.
\]
Also, $\mathbb{P}\phi^*_\theta(Y,Z;\theta_0,\eta_0,F_0)$ is invertible.
\end{condition}

\begin{remark}
Condition 4 presents regularity assumptions on the nuisance parameter estimators $\widehat{\eta}$ and $\widehat{F}$ needed for the following theorem, which states that the asymptotic distribution of the update estimator $\overline{\theta}$ is the same regardless of whether the true values or the estimators of $\eta$ and $F$ are adopted. The first equation in Condition 4 follows if the classes of influence function of the working estimator and its derivatives are Donsker. The second equation requires that substituting the estimators $(\widehat{\eta},\widehat{F})$ for $(\eta_0,F_0)$ does not alter the empirical processes asymptotically, which is an asymptotic equicontinuity requirement. The third equation controls the rate of convergence of $\widehat{\eta}$ and $\widehat{F}$.
\end{remark}

We have the following theorem.

\begin{theorem}
Suppose that $\widehat{\vartheta}_S$ and $\widehat{\vartheta}_F$ are defined as solutions to (\ref{eq:optimal2}) and (\ref{eq:optimal3}), and assume that $\mathrm{E}(\phi^*_0\phi^{*\mathrm{T}}_0)$ is finite and invertible.
Under Conditions 1--5, $n^{1/2}(\overline{\theta}-\theta_0)\rightarrow_d\mathrm{N}(0,\Sigma(\phi^*_0))$, where $\Sigma(\cdot)$ is defined in Theorem 1.
\end{theorem}

\begin{remark}
A key feature of Theorem 2 is that the nuisance parameter estimators need not be root-$n$ consistent. According to Condition 4, we require only a rate of convergence faster than $n^{1/6}$. The reason is that $\widehat{\vartheta}_S$ and $\widehat{\vartheta}_F$ depend on the nuisance parameters through the same functional form, so the leading nuisance-induced errors are identical and cancel in the difference $(\widehat{\vartheta}_S-\widehat{\vartheta}_F)$. 
As a result, the difference, and thus the update estimator, are asymptotically linear.
\end{remark}

\begin{remark}
The cancellation described in Remark 2 does not require \((\widehat\eta,\widehat F)\) to converge to \((\eta_0,F_0)\). Suppose that the same nuisance estimators \((\widehat\eta,\widehat F)\) are used in (\ref{eq:optimal2}) and (\ref{eq:optimal3}) and converge to \((\eta^\dagger,F^\dagger)\). Let \(\vartheta^\dagger\) be the unique solution to $P\phi^*(Y,Z;\vartheta,\eta^\dagger,F^\dagger)=0,$
and set \(\phi^\dagger=\phi^*(Y,Z;\vartheta^\dagger,\eta^\dagger,F^\dagger)\). If the moment conditions in Condition 1 hold for \(\phi^\dagger\), and analogues of Conditions 3–5 hold for \((\vartheta^\dagger,\eta^\dagger,F^\dagger)\), then the proof of Theorem 2 in the Appendix yields $n^{1/2}(\overline\theta-\theta_0)\ \rightarrow_d\
N\{0,\Sigma(\phi^\dagger)\}$. In particular, if $F$ is modeled parametrically and its parameter estimator is root-$n$ consistent for a possibly pseudo-true value, then the requirements concerning $F$ in the analogue of Condition 4 typically follow from standard smoothness and moment conditions. In this case, the update estimator remains valid but is not necessarily optimal.

\end{remark}

\renewcommand{\thecondition}{4\Alph{condition}}
\setcounter{condition}{0}

For a low-dimensional $Z$, we could estimate $F$ nonparametrically using, for example, kernel estimation. In particular, for a $p$-dimensional $Z$, we can set
\begin{align}
\mathrm{d}\widehat{F}(x\mid Z)=\frac{\sum_{j=1}^nR_j\pi_j^{-1}I(X_j=x)\prod_{k=1}^pK\{(Z_{jk}-Z_{k})/a_n\}}{\sum_{j=1}^nR_j\pi_j^{-1}\prod_{k=1}^pK\{(Z_{jk}-Z_k)/a_n\}},\label{eq:kernel}
\end{align}
where $Z_{jk}$ and $Z_{k}$ are the $k$th component of $Z_j$ and $Z$, respectively, $K(\cdot)$ is a density function, and $a_n$ is the bandwidth parameter. Here, we develop the theory for the above kernel estimator for $F$. For simplicity, we focus on $F$ and assume the absence of the nuisance parameter $\eta$. Let $(\mathcal{Y},\mathcal{X},\mathcal{Z})$ be the support of $(Y,X,Z)$. Consider the following conditions.
\begin{condition} The bandwidth parameter $a_n$ satisfies $n^{1/12}a_n\rightarrow0$ and $n^{1/(2p)}a_n\rightarrow\infty$. The density $K$ is symmetric around 0 with $\int z^2K(z)\,\mathrm{d}z<\infty$.
\end{condition}
\begin{condition}
The functions $\psi(y,x,z;\theta)$ and $f(y\mid x,z;\theta)$ are three-times differentiable with respect to $\theta$, with uniformly bounded derivatives. Also, $f(y\mid x,z;\theta)$ and $\mathrm{d}F(x\mid z)$ are uniformly bounded away from 0 over $(y,x,z)\in\mathcal{Y}\times\mathcal{X}\times\mathcal{Z}$.
\end{condition}
\begin{condition}
The covariate $Z$ is continuous, and $\mathcal{Z}$ is bounded. The functions $f_Z(z)$ and $\mathrm{E}(\psi^{(j)}(y,\allowbreak X,\allowbreak z';\theta)f^{(k)}(y\allowbreak\mid X,\allowbreak z';\theta)\allowbreak\mid Z=z)$ are two-times differentiable with uniformly bounded derivatives with respect to $z$ for any $(y,\allowbreak z')\in\mathcal{Y}\times\mathcal{Z}$, where $f_Z$ is the density of $Z$, $\psi^{(j)}$ is the $j$th derivative of $\psi$ with respect to $\theta$, and $f^{(k)}$ is the $k$th derivative of $f$ with respect to $\theta$, for $j,\allowbreak k=0,\allowbreak 1,\allowbreak 2$. Also, $P(R=1\allowbreak\mid Y=y,\allowbreak Z=z)$ is uniformly bounded away from 0 over $(y,\allowbreak z)\in\mathcal{Y}\times\mathcal{Z}$.
\end{condition}

\begin{proposition}
Under Conditions 4A--4C, Condition 4 holds with $\widehat{F}$ defined according to (\ref{eq:kernel}).
\end{proposition}

The proof of Proposition 1 is given in the Appendix.

\begin{remark}
As pointed out in Remark 2, we only require the nuisance parameter estimators to converge at a rate faster than $n^{1/6}$. As a result, Proposition 1 does not require the usual under-smoothing condition for kernel estimation, such as $n^{1/2}a_n^2\rightarrow0$, but only requires a much weaker condition of $n^{1/12}a_n\rightarrow0$. The conditions on $a_n$ imply a constraint that the number of covariates is at most 5. This constraint can be relaxed if we impose stronger smoothness conditions on $\phi^*$ in Condition 4, which would allow a weaker requirement on $r_n$. In practice, however, we do not recommend a fully nonparametric kernel approach when the number of covariates exceeds about 5, even if the theory permits it.
\end{remark}

\subsection{Sequential Update and Joint Update}

When the dimension of $Z$ is moderately high, it is infeasible to fit a fully nonparametric model for $X$ on $Z$. In this case, we could consider different parametric or semiparametric models for $X$, which result in multiple choices of $\widehat{\vartheta}_S$ and $\widehat{\vartheta}_F$. In this subsection, we study how to utilize these estimators to yield a more efficient update estimator $\overline{\theta}$.

Let $(\widehat{\vartheta}_{S,1},\ldots,\widehat{\vartheta}_{S,q})$
be $q$ estimators computed using the subsample and $(\widehat{\vartheta}_{F,1},\ldots,\widehat{\vartheta}_{F,q})$ be the corresponding estimators based on the full sample. These could be estimators with the influence functions (\ref{eq:optimal2}) and (\ref{eq:optimal3}) resulted from different choices of $F$ or estimators from different working models of $Y$ on $Z$. One natural approach is to update the estimator of $\theta$ sequentially, with
\[
\widehat{\theta}^{\mathrm{(Seq)}}_k=\widehat{\theta}^{\mathrm{(Seq)}}_{k-1}-\Sigma_{12,k}^{\mathrm{(Seq)}}\Sigma_{22,k}^{-1}(\widehat{\vartheta}_{S,k}-\widehat{\vartheta}_{F,k})
\]
for $k=1,\ldots,q$, where $\widehat{\theta}^{\mathrm{(Seq)}}_0=\widehat{\theta}_S$, $\Sigma_{12,k}^{\mathrm{(Seq)}}$ is the asymptotic covariance between $n^{1/2}(\widehat{\theta}^{\mathrm{(Seq)}}_{k-1}-\theta_0)$ and $n^{1/2}(\widehat{\vartheta}_{S,k}-\widehat{\vartheta}_{F,k})$, and $\Sigma_{22,k}$ is the asymptotic variance of $n^{1/2}(\widehat{\vartheta}_{S,k}-\widehat{\vartheta}_{F,k})$. The final sequential update estimator is $\widehat{\theta}^{\mathrm{(Seq)}}\equiv\widehat{\theta}^{\mathrm{(Seq)}}_q$. Clearly, $\widehat{\theta}^{\mathrm{(Seq)}}$ is asymptotically at least as efficient as the original estimator $\widehat{\theta}_S$ as well as the sequential update estimators $\widehat{\theta}^{\mathrm{(Seq)}}_k$ for $k=1,\ldots,q-1$.

Another approach to utilize the working estimators $\widehat{\vartheta}_{S,k}$'s and $\widehat{\vartheta}_{F,k}$'s is to concatenate the estimators to form $\widehat{\vartheta}_S=(\widehat{\vartheta}_{S,1}^{\mathrm{T}},\ldots,\widehat{\vartheta}_{S,q}^{\mathrm{T}})^{\mathrm{T}}$ and $\widehat{\vartheta}_F=(\widehat{\vartheta}_{F,1}^{\mathrm{T}},\ldots,\widehat{\vartheta}_{F,q}^{\mathrm{T}})^{\mathrm{T}}$ and define the joint update estimator
\begin{align*}
\widehat{\theta}^{\mathrm{(Joint)}}=\widehat{\theta}_S-\Sigma_{12}\Sigma_{22}^{-1}(\widehat{\vartheta}_S-\widehat{\vartheta}_F),
\end{align*}
where $\Sigma_{12}$ is the asymptotic covariance between $n^{1/2}(\widehat{\theta}_S-\theta_0)$ and $n^{1/2}(\widehat{\vartheta}_S-\widehat{\vartheta}_F)$, and $\Sigma_{22}$ is the asymptotic variance of $n^{1/2}(\widehat{\vartheta}_S-\widehat{\vartheta}_F)$.

Similar to the above, we assume that each working estimator is asymptotically linear with
\[
n^{1/2}(\widehat{\vartheta}_{S,k}-\vartheta_0) = n^{-1/2} \sum_{i=1}^n \frac{R_i}{\pi_i} \, \phi_{k}(Y_i,Z_i;\theta_0,\eta_0,F_0) + o_p(1)
\]
and
\[
n^{1/2}(\widehat{\vartheta}_{F,k}-\vartheta_0) = n^{-1/2} \sum_{i=1}^n\phi_k(Y_i,Z_i;\theta_0,\eta_0,F_0)+ o_p(1)
\]
for $k=1,\ldots,q$, where $\phi_k(Y,Z;\theta_0,\eta_0,F_0)$ is mean zero. We have the following theorem concerning the relative efficiency of the sequential and joint update estimators.

\begin{theorem}
Let $\Sigma_0$, $\Sigma_{k}^{\mathrm{(Seq)}}$, and $\Sigma^{\mathrm{(Joint)}}$ be the asymptotic covariance matrices of $n^{1/2}(\widehat{\theta}_S-\theta_0)$, $n^{1/2}(\widehat{\theta}^{\mathrm{(Seq)}}_k-\theta_0)$, and $n^{1/2}(\widehat{\theta}^{\mathrm{(Joint)}}-\theta_0)$, respectively, where $k=1,\ldots,q$. Assume Conditions 1 and 2, where Condition 1 holds for $\phi=(\phi_1^{\mathrm{T}},\ldots,\phi_q^{\mathrm{T}})^{\mathrm{T}}$. We have
\[
\Sigma_0\succeq \Sigma_{1}^{\mathrm{(Seq)}}\succeq \cdots\succeq \Sigma_{q}^{\mathrm{(Seq)}}\succeq \Sigma^{\mathrm{(Joint)}}.
\]
\end{theorem}

From Theorem~3, we see that whenever there are multiple working estimators, it is always preferable to concatenate them into a longer vector of estimators and perform the joint update than to use them to update the estimator sequentially. The joint update estimator also avoids the need to order the set of working estimators. In the sequel, whenever we have multiple working estimators, we adopt the joint update approach instead of the sequential update approach.

Note that both the sequential update and joint update estimators are consistent and asymptotically at least as efficient as the original estimator, regardless of the choices of working estimators. Also, an implication of Theorem~3 is that the joint update estimator is optimally efficient if one of the $(\widehat{\vartheta}_{S,k},\widehat{\vartheta}_{F,k})$'s is optimal, and there is no cost (asymptotically) for considering multiple sets of estimators. We propose to set the components of $\widehat{\vartheta}_S$ as the solutions to \eqref{eq:optimal2} based on the subsample with different choices of $\widehat{F}$, and then obtain the components of $\widehat{\vartheta}_F$ correspondingly based on the full sample as the solutions to \eqref{eq:optimal3}. For example, for a more flexible association structure, we can fit a generalized additive model of $X$ against the covariates $Z$, where the effect of each component of $Z$ is modeled nonparametrically. Also, we can include the estimator under a standard working model for $Y$ on $Z$, as typically adopted in the literature, to safeguard against loss of efficiency from the existing methods. 

\subsection{Algorithms and Examples}

The proposed optimal update estimator can be obtained by the following steps: 
\begin{enumerate}
\item[1.] Compute the original estimator $\widehat{\theta}_S$ based on the subsample.
\item[2.] Compute the estimators of the nuisance parameters $\eta$ and $F$, denoted by $\widehat{\eta}$ and $\widehat{F}$, based on the subsample.
\item[3.] Obtain the working estimators $\widehat{\vartheta}_S$ and $\widehat{\vartheta}_F$ by solving \eqref{eq:optimal2} and \eqref{eq:optimal3} based on the subsample and the full sample, respectively.
\item[4.] Compute the estimators of the covariance matrices $\Sigma_{12}$ and $\Sigma_{22}$, denoted by $\widehat\Sigma_{12}$ and $\widehat\Sigma_{22}$. In particular, we compute the influence functions of $\widehat{\theta}_S$ and $(\widehat{\vartheta}_S-\widehat{\vartheta}_F)$ for the $i$th subject as
\[
\frac{R_i}{\pi_i}\psi(Y_i,X_i,Z_i;\widehat{\theta}_S,\widehat{\eta})\quad\text{and}
\quad\bigg(\frac{R_i}{\pi_i}-1\bigg)\bigg\{-\frac{1}{n}\sum_{j=1}^n\phi^*_\theta(Y_j,Z_j;\widehat{\vartheta}_F,\widehat{\eta},\widehat{F})\bigg\}^{-1}\phi^*(Y_i,Z_i;\widehat{\vartheta}_F,\widehat{\eta},\widehat{F}),\]
respectively, for $i=1,\ldots,n$. The matrices $\widehat{\Sigma}_{12}$ and $\widehat{\Sigma}_{22}$ are the corresponding sample covariance matrices of these quantities.
\item[5.] Compute the update estimator $\overline{\theta}$ using \eqref{eq:update} with $\Sigma_{12}$ and $\Sigma_{22}$ replaced by their estimators $\widehat\Sigma_{12}$ and $\widehat\Sigma_{22}$ obtained in Step 4.
\item[6.] Repeat Steps 1--5 on $B$ bootstrap samples and obtain the estimator of the covariance matrix of $\overline{\theta}$ using the sample covariance matrix of the bootstrap estimators $\{\overline{\theta}^{(b)}, b=1,\ldots,B\}$.
\end{enumerate}

The joint update estimator can be calculated by similar steps as described above, except that the working estimators $\widehat{\vartheta}_S$ and $\widehat{\vartheta}_F$ in Step 3 may consist of multiple components obtained under different choices of $\widehat{F}$ or under different working models of $Y$ on $Z$.

\vspace{0.1in}

Here are some remarks on the implementation of the proposed update estimators:
\begin{enumerate}
\item[a.]In Step 3, we suggest to solve \eqref{eq:optimal2} and \eqref{eq:optimal3} using the Broyden or Newton method as implemented in, for example, the R function \texttt{nleqslv}. The proposed kernel estimator $\widehat{F}(\cdot \mid Z)$ given in (\ref{eq:kernel}) is a weighted discrete distribution supported on the observed $X_j$'s. Consequently, the integrals in \eqref{eq:optimal2} and \eqref{eq:optimal3} reduce to finite weighted sums obtained by evaluating the integrands at the observed $X_j$'s, regardless of $X$ being one- or multi-dimensional. Alternatively, one may specify a parametric model for $X$ given $Z$. Under a regression model with Gaussian errors, the conditional mean of $X$ can be estimated under a specified structure, such as a linear or generalized additive structure, and the error variance can be estimated from the residuals. When $X$ is one-dimensional, the integrals in \eqref{eq:optimal2} and \eqref{eq:optimal3} can be approximated using Gauss--Hermite quadrature. When $X$ is multi-dimensional, tensor-product Gauss--Hermite quadrature can be used. In this case, if the integrands depend on $X$ only through a linear predictor, then the transformation technique of \cite{kwok2026computationally} can be adopted to reduce the required numerical integration to one dimension.
\item[b.]In some cases, the working estimators $\widehat{\vartheta}_S$ and $\widehat{\vartheta}_F$ may be too extreme due to numerical issues. In such cases, we suggest to add a penalty term to \eqref{eq:optimal2} and \eqref{eq:optimal3} to regularize the working estimators. For example, we can employ the $L_2$ penalty by adding the terms $-2\lambda n_0^{-1/3} \theta$ and $-2\lambda n^{-1/3} \theta$ to \eqref{eq:optimal2} and \eqref{eq:optimal3}, with $n_0$ being the size of the Phase II sample. To select the tuning parameter $\lambda$, we can consider a set of small values and select the one that does not yield any outliers in $B$ bootstraps. %
\item[c.] For the joint update estimator, the multiple sets of working estimators $\widehat{\vartheta}_S$ and $\widehat{\vartheta}_F$ could be highly correlated, leading to a nearly singular covariance matrix $\widehat\Sigma_{22}$. To avoid numerical issues, in Step 4, we could directly estimate $\Sigma_{12}\Sigma_{22}^{-1}$ using ridge regression instead. Specifically, we take the influence function of $\widehat{\theta}_S$ as the outcome and take the influence function of $(\widehat{\vartheta}_S-\widehat{\vartheta}_F)$ as the covariates in a multivariate linear ridge regression. The estimated regression coefficient is then used as an estimate of $\Sigma_{12}\Sigma_{22}^{-1}$. The ridge regression can be implemented using the R function \texttt{glmnet} with the \texttt{mgaussian} family. The tuning parameter can be selected by cross-validation using the R function \texttt{cv.glmnet}. If the tuning parameter is set to be zero, then the estimate reduces to  $\widehat\Sigma_{12}\widehat\Sigma_{22}^{-1}$ given in Step 4.
\item[d.] There is an alternative, more computationally efficient way to implement the proposed method by using the same bootstrap samples to obtain the update term and estimate the variance of the update estimator. Specifically, in Step 4, we repeat Steps 1--3 on $B$ bootstrap samples and obtain the estimates $\widehat\Sigma_{11}$, $\widehat\Sigma_{12}$, $\widehat\Sigma_{21}$ and $\widehat\Sigma_{22}$ using the sample covariance matrix of the bootstrap estimators $\{(\widehat{\theta}_S^{(b)},\widehat{\vartheta}_S^{(b)}-\widehat{\vartheta}_F^{(b)}), b=1,\ldots,B\}$. In Step 6, we do not perform another round of bootstrap but simply estimate the covariance matrix of $\overline{\theta}$ by $n^{-1}(\widehat\Sigma_{11}-\widehat\Sigma_{12}\widehat\Sigma_{22}^{-1}\widehat\Sigma_{21})$. This algorithm tends to slightly underestimate the variance of $\overline{\theta}$ especially under MAR, but it has a lower computational cost. We refer to this method as the fully bootstrap method.
\end{enumerate}

We illustrate the proposed methods with three common models. First, consider the linear model 
$$
Y = W^{\mathrm{T}}\theta + \epsilon, \quad \epsilon\sim N(0,\sigma^2),
$$
where $W=(X^{\mathrm{T}},Z^{\mathrm{T}})^{\mathrm{T}}$ is the vector of covariates, and $\theta$ is the vector of regression coefficients. The influence function for the least-squares estimator of $\theta$ is proportional to 
$$
\psi(Y,X,Z)=(Y-W^{\mathrm{T}}\theta)W.
$$
With this $\psi$, we can perform the optimal or joint update estimation of $\theta$ following the steps given in the above algorithm, where the nuisance parameter $\sigma$ can be estimated by the MLE or IPW based on the subsample.

Second, consider the logistic model 
$$
P(Y=y|X,Z) = \frac{e^{yW^{\mathrm{T}}\theta}}{1+e^{W^{\mathrm{T}}\theta}}
$$
for $y=0$ and 1, where $\theta$ is the vector of regression coefficients. The influence function for the MLE of $\theta$ is proportional to 
$$
\psi(Y,X,Z)=\big\{Y-(1+e^{-W^{\mathrm{T}}\theta})^{-1}\big\}W.
$$
The optimal or joint update estimator of $\theta$ can be obtained by the steps given above.

Lastly, consider the Cox model with an event time of interest $\widetilde{T}$. Under the Cox model, the cumulative hazard function of $\widetilde{T}$ conditional on $X$ and $Z$ is given by
\[
\Lambda(t | X,Z) = \Lambda(t) \exp(W^{\mathrm{T}}\theta),
\]
where $\Lambda(\cdot)$ is an unspecified baseline cumulative hazard function, and $\theta$ is the vector of regression coefficients. In practice, $\widetilde{T}$ is often subject to right censoring. Let $C$ denote the censoring time, $T\equiv\min\{\widetilde{T},C\}$ denote the observed time, and $\Delta\equiv I(\widetilde{T}\leq C)$ denote the event indicator. The influence function for the maximum partial likelihood estimator of $\theta$ is proportional to 
\begin{align*}
\psi(Y,X,Z)=\int \bigg\{W-\frac{s^{(1)}(\theta,t)}{s^{(0)}(\theta,t)}\bigg\} \,\mathrm{d}\mathcal{N}(t)-\int \frac{\mathcal{Y}(t)e^{W^{\mathrm{T}}\theta}}{s^{(0)}(\theta,t)}\bigg\{W-\frac{s^{(1)}(\theta,t)}{s^{(0)}(\theta,t)}\bigg\}\,\mathrm{d}F(t),
\end{align*}
where $\mathcal{Y}(t)=I(T\geq t)$, $s^{(r)}(\theta,t)=\mathrm{E}\{\mathcal{Y}(t)e^{W^{\mathrm{T}}\theta}W^{\otimes r}\}$ for $r=0,1$, $\mathcal{N}(t)=I(T\leq t, \Delta=1)$, $F(t)=\mathrm{E}\{\mathcal{N}(t)\}$, $W^{\otimes0}=1$, and $W^{\otimes1}=W$ \citep{lin1989robust}. 
With this $\psi$, we can perform the optimal or joint update following the steps given in the above algorithm, where $s^{(r)}$ and $F$ in $\psi$ can be replaced by the empirical estimators based on the subsample and the full sample, respectively, and the nuisance parameter $\Lambda$ can be estimated by the (weighted) nonparametric MLE from the subsample.

\section{Simulation Studies}

We conduct three sets of simulation studies to demonstrate the feasibility and superiority of the proposed methods under some common models. The first two sets of simulations are conducted under MCAR, while the last one is under MAR. Specifically, in the first simulation study, we evaluate the performance of the optimal update estimator under three common models:
\begin{enumerate}
\item[S1.] The linear model: $Y=\theta X + \epsilon$ with $\theta=0.5$ and $\epsilon\sim N(0,0.25)$; 
\item[S2.] The logistic model: $P(Y=1|X)=1/(1+e^{-\theta X})$ with $\theta=0.5$; 
\item[S3.] The Cox model: $\lambda(t\mid X) = \lambda(t) e^{\theta X}$ with $\lambda(t)=0.5t$ and $\theta=\log 2$. 
\end{enumerate}
For the Cox model, the survival outcome $Y$ consists of $T=\min\{\widetilde{T},C\}$ and $\Delta=I(\widetilde{T}\leq C)$, where $\widetilde{T}$ is the failure time, and $C$ is the censoring time. We generate $\widetilde{T}$ from the Cox model given in S3 and generate $C$ from $\min\{\mathrm{Unif}(0,5\tau/3),\tau\}$ with $\tau$ chosen to yield a censoring rate of about $50\%$. We consider two settings for the covariate: (i) $X$ is continuous and follows $N(0,1)$; (ii) $X$ is discrete and follows $\mathrm{Ber}(0.5)$. For continuous $X$, we generate an auxiliary variable $X^*=X+e$, where $e\sim N(0,\sigma^2_e)$ with $\sigma_e$ chosen such that $\rho \equiv \mathrm{Corr}(X,X^*)=0.3$ or $0.7$. For discrete $X$, we generate $X^*$ as a binary auxiliary variable for $X$ with misclassification rate $\rho=0.1$ or $0.3$. For each scenario, we consider three estimators for comparison: (i) the original estimator based on the subsample only; (ii) the update estimator using the default working outcome model, i.e., the linear, logistic, or Cox model of $Y$ on $X^*$; and (iii) the update estimator using the proposed optimal update procedure. For continuous $X$, we use the kernel estimator given by \eqref{eq:kernel} to estimate the conditional distribution of $X$ given $X^*$, with the bandwidth chosen according to Silverman's rule of thumb \citep{silverman1986density}. For discrete $X$, we use the empirical estimator of the conditional distribution of $X$ given $X^*$. The full sample size is $n=1000$, and the subsample is randomly selected under MCAR with a size of $400$. For the logistic model, we notice that the working estimators have some outliers, and thus add the $L_2$ penalty to \eqref{eq:optimal2} and \eqref{eq:optimal3} to regularize the working estimators as described in Section 2.4. For the tuning parameter $\lambda$, we consider the values 0.005, 0.01, 0.02, and 0.04. Similarly, we add a penalty term for the logistic model in subsequent simulation studies.

The simulation results based on $500$ replicates are presented in Table~\ref{tab1}. One can see from the results that for all scenarios considered, the three methods are unbiased and yield accurate standard error estimation with desired coverage. Both the optimal and default update estimators are more efficient than the original estimator, and the efficiency gain increases as the correlation $\rho$ between $X$ and $X^*$ increases in the continuous setting or as the misclassification rate $\rho$ between $X$ and $X^*$ decreases in the discrete setting. The proposed optimal update estimator is generally more efficient than the default update estimator, with substantial improvement in some cases.

\begin{table}\small
    \tbl{Simulation results with one covariate (MCAR)}
	{
    \renewcommand{\tabcolsep}{2bp}
    \renewcommand{\arraystretch}{1.2}
    \begin{tabular}{clrcc cccrc ccccr cccc}
		\hline
		&    &    \multicolumn{5}{c}{Linear Model}  & &   \multicolumn{5}{c}{Logistic Model}    &  &     \multicolumn{5}{c}{Cox Model}    \\ \cline{3-7}\cline{9-13}\cline{15-19}
		$\rho$ & Method   &     Bias &  SSD  &  ESE  & CP   &  RE  &  &     Bias &  SSD  & ESE   &  CP  &  RE  &  &     Bias &  SSD  & ESE   &  CP  &  RE \\ \hline
        &                 &        \multicolumn{17}{c}{Continuous Covariate} \\
		0.3   & Original &   $-$0.001 &0.024 &0.025 &0.962 &1.000 & & $-$0.005 &0.106 &0.110 &0.952 &1.000 &  &  0.009 &0.079 &0.079 &0.954 &1.000  \\
		& Default  &   $-$0.001 &0.024 &0.025 &0.944 &1.041  & & $-$0.005 &0.102 &0.107 &0.952 &1.072 &  &  0.009 &0.077 &0.078 &0.956 &1.055  \\
		& Optimal  & $-$0.001 &0.021 &0.021 &0.952 &1.377 & & $-$0.013 &0.101 &0.106 &0.958 &1.096 &  & 0.010 &0.074 &0.079 &0.966 &1.139  \\
		0.7   & Original &  $-$0.001 &0.024 &0.025 &0.962 &1.000 & & $-$0.005 &0.106 &0.110 &0.954 &1.000 &  & 0.009 &0.079 &0.079 &0.954 &1.000  \\
		& Default  &  0.000 &0.022 &0.022 &0.948 &1.268 & & $-$0.005 &0.087 &0.094 &0.960 &1.476 &  & 0.009 &0.070 &0.072 &0.956 &1.279  \\
		& Optimal  & 0.000 &0.020 &0.021 &0.952 &1.485 & & $-$0.010 &0.087 &0.092 &0.958 &1.503 &  &  0.008 &0.068 &0.070 &0.970 &1.364 \\ \hline
        &                 &        \multicolumn{17}{c}{Discrete Covariate} \\
		0.1   & Original &    0.000 &0.037 &0.035 &0.936 &1.000 & & $-$0.005 &0.144 &0.146 &0.956 &1.000 &  &  0.010 &0.146 &0.148 &0.944 &1.000  \\
		& Default  &    0.001 &0.028 &0.026 &0.936 &1.772  & & $-$0.002 &0.104 &0.105 &0.952 &1.938 &  &  0.006 &0.118 &0.116 &0.952 &1.544  \\
		& Optimal  & 0.001 &0.027 &0.026 &0.940 &1.890 & & $-$0.006 &0.102 &0.104 &0.948 &1.995 &  & 0.004 &0.118 &0.118 &0.958 &1.532  \\
		0.3   & Original &   0.000 &0.037 &0.035 &0.936 &1.000 & & $-$0.005 &0.144 &0.146 &0.954 &1.000 &  & 0.010 &0.146 &0.148 &0.944 &1.000  \\
		& Default  &  0.000 &0.032 &0.031 &0.938 &1.297 & & $-$0.008 &0.122 &0.124 &0.960 &1.391 &  & 0.007 &0.139 &0.140 &0.934 &1.109  \\
		& Optimal  & 0.001 &0.030 &0.028 &0.938 &1.542 & & $-$0.011 &0.116 &0.118 &0.960 &1.553 &  &  0.011 &0.142 &0.150 &0.956 &1.061 \\ \hline
	\end{tabular}}
    \begin{tabnote}
    NOTE: Bias is the average point estimate minus the true value, SSD is the sample standard deviation, ESE is the average estimated standard error based on $100$ bootstrap samples, CP is the coverage proportion of the $95\%$ confidence interval based on the normal approximation, and RE is the relative efficiency with respect to the original estimator
    \end{tabnote}
	\label{tab1}
\end{table}

In the second simulation study, we evaluate the performance of the joint update estimator under the following models: 
\begin{enumerate}
\item[M1.] The linear model: $Y=\beta_1 X + \beta_2 Z_1 + \beta_3 Z_2 + \beta_4 Z_3 + \beta_5 Z_4 + \epsilon$ with $\beta=(1,0.5,-0.5,0,0)$ and $\epsilon\sim N(0,0.25)$; 
\item[M2.] The logistic model: $P(Y=1|X,Z_1,Z_2,Z_3,Z_4)=1/(1+\exp\{-(\beta_1 X + \beta_2 Z_1 + \beta_3 Z_2 + \beta_4 Z_3 + \beta_5 Z_4)\})$ with $\beta=(1,0.5,-0.5,0,0)$;
\item[M3.] The Cox model: $\lambda(t\mid X,Z_1,Z_2,Z_3,Z_4) = \lambda(t) \exp\{\beta_1 X + \beta_2 Z_1 + \beta_3 Z_2 + \beta_4 Z_3 + \beta_5 Z_4\}$ with $\lambda(t)=0.5t$ and $\beta=(\log 2,0.5,-0.5,0,0)$.
\end{enumerate}
For each model, we consider two settings for the covariates as follows. Under Covariate Setting I, $(Z_1,Z_2,Z_3,Z_4)$ are independent normal random variables with mean 0 and variance 1, and $X=\theta(Z_1+Z_2+Z_3+Z_4)+\varepsilon$ with $\varepsilon\sim N(0,\sigma^2_\varepsilon)$ and $(\theta,\sigma_\varepsilon)$ chosen such that the coefficient of determination $R^2=0.7$. We generate the auxiliary variable $X^*=X+e$ with $e\sim N(0,\sigma^2_e)$, where $\sigma_e$ is chosen such that $\rho \equiv \mathrm{Corr}(X,X^*)=0.3$ or $0.7$. Under Covariate Setting II, $(\log Z_1, \log Z_2, \log Z_3, \log Z_4)$ are independent truncated normal random variables with mean 0, variance 1, and truncated by 2 to avoid extreme observations. We generate $X=\theta(\log Z_1 + \log Z_2 + \log Z_3 + \log Z_4)+\varepsilon$ with $\varepsilon\sim N(0,\sigma^2_\varepsilon)$ and adjust $(\theta,\sigma_\varepsilon)$ such that $X$ and $(Z_1,Z_2,Z_3,Z_4)$ have $R^2=0.7$. Mimicking the case with detection limit, we generate the auxiliary variable as $X^*=I(X\leq -1)(-1)+I(X>-1)X+e$, where $e\sim N(0,\sigma^2)$ with $\sigma$ chosen such that $\rho \equiv \mathrm{Corr}(X,X^*)=0.3$ or $0.7$. With a slight abuse of notation, in the sequel, we use $Z$ to denote $(Z_1,Z_2,Z_3,Z_4)^{\mathrm{T}}$, although $Z$ in the methodological development would include the auxiliary variable $X^*$ as well.

We compare the following seven estimators: (i) the original estimator based on the subsample only; (ii) the update estimator using the default working outcome model, i.e., the linear, logistic, or Cox model of $Y$ on $(X^*,Z)$; (iii) the update estimator based on \eqref{eq:optimal2} using the linear working model of $X$ on $(X^*,Z)$; (iv) the joint update estimator that combines the default update estimator (ii) and the linear update estimator (iii); (v) the update estimator based on \eqref{eq:optimal2} using the generalized additive model (GAM) as the working model of $X$ on $(X^*,Z)$; (vi) the joint update estimator that combines the default update estimator (ii) and the GAM update estimator (v); (vii) the optimal update estimator assuming known distribution of $(X,X^*,Z)$. For the update estimators, we use ridge regression in Step 4 as described in Section 2.4, where the tuning parameter is selected by cross-validation. The full sample size is $n=1000$ and the subsample is randomly selected under MCAR with a size of $400$.

The simulation results based on $500$ replicates are given in Tables~\ref{tab2}, \ref{stab1}, and \ref{stab2} for the linear model, the logistic model, and the Cox model, respectively. Only the results for the first two regression coefficients are presented due to the limited space. One can see that under both covariate settings, the seven methods are unbiased and yield accurate standard error estimation with desired coverage. The update estimators are more efficient than the original estimator, and the efficiency gain increases with the correlation $\rho$ between $X$ and $X^*$. For Covariate Setting I, the default update estimator is generally the least efficient among all update estimators. As expected, the linear and GAM update estimators perform as well as the optimal update estimator, since the linear and GAM models for $X$ on $(X^*,Z)$ include the true model. The two joint update estimators are also as good as the optimal update estimator as expected. For Covariate Setting II, since the linear model of $X$ on $(X^*,Z)$ is very different from the true model, the linear update performs poorly. Nevertheless, its joint update with the default performs at least as well as the default, which shows the advantage of using joint update to safeguard against loss of efficiency. The GAM update estimator in this setting still performs comparably to the optimal update estimator, since the GAM model can capture the non-linear effects of $(X^*,Z)$ on $X$.

\begin{table}\small
	\tbl{Simulation results with multiple covariates under the linear model (MCAR)}
	{
        \renewcommand{\tabcolsep}{4bp}
    \renewcommand{\arraystretch}{1.2}
\begin{tabular}{clrcc cccrc ccc}
		\hline
		&                 &     \multicolumn{5}{c}{$\beta_1=1$}      &  &    \multicolumn{5}{c}{$\beta_2=0.5$}    \\ \cline{3-7}\cline{9-13}
		$\rho$ & Method          &     Bias &  SSD  &  ESE  & CP   &  RE  &  &     Bias &  SSD  & ESE   &  CP  &  RE  \\ \hline
            &                 &        \multicolumn{11}{c}{Covariate Setting I} \\
  0.3   & Original        & 0.001 & 0.039 & 0.039 & 0.950 & 1.000 && $-$0.000 & 0.032 & 0.032 & 0.962 & 1.000 \\ 
  & Default         & $-$0.000 & 0.039 & 0.039 & 0.956 & 1.008 && 0.000 & 0.030 & 0.030 & 0.948 & 1.109 \\ 
  & Linear          & 0.005 & 0.034 & 0.033 & 0.928 & 1.340 && $-$0.003 & 0.029 & 0.028 & 0.944 & 1.256 \\ 
  & Default+Linear  & 0.004 & 0.034 & 0.033 & 0.930 & 1.343 && $-$0.002 & 0.029 & 0.028 & 0.946 & 1.256 \\ 
  & GAM             & 0.007 & 0.034 & 0.034 & 0.930 & 1.325 && $-$0.003 & 0.029 & 0.028 & 0.950 & 1.254 \\ 
  & Default+GAM     & 0.006 & 0.034 & 0.034 & 0.934 & 1.327 && $-$0.003 & 0.029 & 0.028 & 0.944 & 1.252 \\ 
  & Optimal         & 0.005 & 0.034 & 0.033 & 0.926 & 1.343 && $-$0.003 & 0.029 & 0.028 & 0.942 & 1.261 \\ 
  0.7   & Original        & 0.001 & 0.039 & 0.039 & 0.950 & 1.000 && $-$0.000 & 0.032 & 0.032 & 0.962 & 1.000 \\ 
  & Default         & 0.000 & 0.038 & 0.038 & 0.958 & 1.076 && $-$0.000 & 0.030 & 0.029 & 0.948 & 1.164 \\ 
  & Linear          & 0.005 & 0.034 & 0.033 & 0.924 & 1.350 && $-$0.003 & 0.028 & 0.027 & 0.940 & 1.298 \\ 
  & Default+Linear  & 0.004 & 0.034 & 0.033 & 0.926 & 1.346 && $-$0.002 & 0.028 & 0.028 & 0.940 & 1.298 \\ 
  & GAM             & 0.005 & 0.034 & 0.034 & 0.934 & 1.334 && $-$0.003 & 0.028 & 0.028 & 0.940 & 1.302 \\ 
  & Default+GAM     & 0.005 & 0.034 & 0.034 & 0.938 & 1.336 && $-$0.003 & 0.028 & 0.028 & 0.944 & 1.302 \\ 
  & Optimal         & 0.005 & 0.034 & 0.033 & 0.926 & 1.352 && $-$0.003 & 0.028 & 0.027 & 0.944 & 1.305 \\ \hline
            &                 &        \multicolumn{11}{c}{Covariate Setting II} \\
  0.3   & Original        & -0.001 & 0.015 & 0.015 & 0.946 & 1.000 && 0.001 & 0.017 & 0.017 & 0.952 & 1.000 \\ 
  & Default         & $-$0.001 & 0.015 & 0.015 & 0.948 & 0.997 && 0.001 & 0.016 & 0.016 & 0.956 & 1.091 \\ 
  & Linear          & 0.000 & 0.014 & 0.014 & 0.948 & 1.109 && 0.001 & 0.015 & 0.016 & 0.954 & 1.174 \\ 
  & Default+Linear  & 0.000 & 0.014 & 0.014 & 0.948 & 1.116 && 0.001 & 0.015 & 0.016 & 0.948 & 1.185 \\ 
  & GAM             & 0.001 & 0.014 & 0.014 & 0.950 & 1.251 && 0.001 & 0.014 & 0.014 & 0.952 & 1.427 \\ 
  & Default+GAM     & 0.001 & 0.014 & 0.014 & 0.948 & 1.251 && 0.001 & 0.014 & 0.014 & 0.964 & 1.431 \\ 
  & Optimal         & $-$0.000 & 0.013 & 0.013 & 0.962 & 1.417 && 0.001 & 0.014 & 0.014 & 0.956 & 1.509 \\ 
  0.7   & Original        & -0.001 & 0.015 & 0.015 & 0.946 & 1.000 && 0.001 & 0.017 & 0.017 & 0.952 & 1.000 \\ 
  & Default         & $-$0.001 & 0.015 & 0.015 & 0.938 & 1.024 && 0.001 & 0.016 & 0.016 & 0.952 & 1.143 \\ 
  & Linear          & 0.000 & 0.014 & 0.015 & 0.950 & 1.104 && 0.001 & 0.015 & 0.016 & 0.958 & 1.216 \\ 
  & Default+Linear  & $-$0.000 & 0.014 & 0.014 & 0.952 & 1.112 && 0.001 & 0.015 & 0.016 & 0.950 & 1.235 \\ 
  & GAM             & 0.001 & 0.014 & 0.014 & 0.958 & 1.244 && 0.001 & 0.014 & 0.014 & 0.952 & 1.454 \\ 
  & Default+GAM     & 0.001 & 0.014 & 0.014 & 0.956 & 1.252 && 0.001 & 0.014 & 0.014 & 0.956 & 1.452 \\ 
  & Optimal         & $-$0.000 & 0.013 & 0.013 & 0.962 & 1.439 && 0.002 & 0.013 & 0.014 & 0.960 & 1.541 \\ \hline
	\end{tabular}}
    \begin{tabnote}
    NOTE: See NOTE to Table \ref{tab1}
    \end{tabnote}
	\label{tab2}
\end{table}

\begin{table}[htbp]\small
\tbl{Simulation results with multiple covariates under the logistic model (MCAR)}
	{
        \renewcommand{\tabcolsep}{4bp}
    \renewcommand{\arraystretch}{1.2}
\begin{tabular}{clrcc cccrc ccc}
		\hline
		&                 &     \multicolumn{5}{c}{$\beta_1=1$}      &  &    \multicolumn{5}{c}{$\beta_2=0.5$}    \\ \cline{3-7}\cline{9-13}
		$\rho$ & Method          &     Bias &  SSD  &  ESE  & CP   &  RE  &  &     Bias &  SSD  & ESE   &  CP  &  RE  \\ \hline
  &                 &        \multicolumn{11}{c}{Covariate Setting I} \\
  0.3   & Original        & 0.013 & 0.188 & 0.203 & 0.966 & 1.000 && 0.015 & 0.150 & 0.160 & 0.958 & 1.000 \\ 
  & Default         & 0.013 & 0.188 & 0.203 & 0.956 & 0.996 && 0.003 & 0.114 & 0.117 & 0.964 & 1.727 \\ 
  & Linear          & 0.008 & 0.187 & 0.202 & 0.962 & 1.011 && $-$0.012 & 0.113 & 0.121 & 0.964 & 1.760 \\ 
  & Default+Linear  & 0.012 & 0.192 & 0.210 & 0.970 & 0.961 && 0.004 & 0.116 & 0.124 & 0.964 & 1.671 \\ 
  & GAM             & $-$0.010 & 0.186 & 0.196 & 0.952 & 1.024 && $-$0.010 & 0.113 & 0.120 & 0.966 & 1.741 \\ 
  & Default+GAM     & $-$0.011 & 0.187 & 0.201 & 0.958 & 1.009 && 0.011 & 0.114 & 0.120 & 0.966 & 1.720 \\ 
  & Optimal         & 0.010 & 0.187 & 0.202 & 0.960 & 1.011 && $-$0.013 & 0.113 & 0.122 & 0.964 & 1.750 \\ 
  0.7   & Original        & 0.013 & 0.188 & 0.203 & 0.962 & 1.000 && 0.015 & 0.150 & 0.160 & 0.968 & 1.000 \\ 
  & Default         & 0.010 & 0.182 & 0.193 & 0.958 & 1.066 && 0.003 & 0.113 & 0.115 & 0.944 & 1.749 \\ 
  & Linear          & $-$0.003 & 0.180 & 0.192 & 0.956 & 1.086 && $-$0.000 & 0.111 & 0.114 & 0.946 & 1.816 \\ 
  & Default+Linear  & 0.003 & 0.182 & 0.196 & 0.962 & 1.070 && 0.007 & 0.114 & 0.119 & 0.950 & 1.711 \\ 
  & GAM             & $-$0.017 & 0.178 & 0.185 & 0.958 & 1.108 && 0.005 & 0.111 & 0.112 & 0.948 & 1.822 \\ 
  & Default+GAM     & $-$0.021 & 0.180 & 0.188 & 0.946 & 1.095 && 0.015 & 0.112 & 0.115 & 0.950 & 1.785 \\ 
  & Optimal         & $-$0.001 & 0.180 & 0.192 & 0.956 & 1.086 && $-$0.001 & 0.112 & 0.114 & 0.944 & 1.803 \\ \hline
  &                 &        \multicolumn{11}{c}{Covariate Setting II} \\
  0.3   & Original        & 0.027 & 0.114 & 0.119 & 0.960 & 1.000 && 0.020 & 0.109 & 0.112 & 0.976 & 1.000 \\ 
  & Default         & 0.026 & 0.113 & 0.118 & 0.960 & 1.010 && 0.010 & 0.083 & 0.083 & 0.960 & 1.709 \\ 
  & Linear          & 0.002 & 0.091 & 0.097 & 0.976 & 1.567 && 0.001 & 0.076 & 0.076 & 0.962 & 2.068 \\ 
  & Default+Linear  & 0.005 & 0.091 & 0.097 & 0.974 & 1.553 && 0.004 & 0.077 & 0.076 & 0.960 & 2.012 \\ 
  & GAM             & $-$0.004 & 0.080 & 0.082 & 0.950 & 2.022 && $-$0.002 & 0.072 & 0.070 & 0.954 & 2.307 \\ 
  & Default+GAM     & $-$0.003 & 0.080 & 0.083 & 0.948 & 2.029 && $-$0.000 & 0.072 & 0.071 & 0.954 & 2.297 \\ 
  & Optimal         & $-$0.001 & 0.080 & 0.081 & 0.958 & 2.044 && $-$0.000 & 0.071 & 0.070 & 0.952 & 2.327 \\ 
  0.7   & Original        & 0.027 & 0.114 & 0.119 & 0.962 & 1.000 && 0.020 & 0.109 & 0.112 & 0.976 & 1.000 \\ 
  & Default         & 0.020 & 0.103 & 0.106 & 0.950 & 1.211 && 0.009 & 0.078 & 0.079 & 0.962 & 1.943 \\ 
  & Linear          & 0.002 & 0.088 & 0.093 & 0.964 & 1.682 && 0.000 & 0.074 & 0.074 & 0.956 & 2.169 \\ 
  & Default+Linear  & 0.007 & 0.089 & 0.094 & 0.958 & 1.625 && 0.004 & 0.075 & 0.075 & 0.964 & 2.115 \\ 
  & GAM             & $-$0.004 & 0.079 & 0.081 & 0.942 & 2.062 && $-$0.001 & 0.071 & 0.070 & 0.946 & 2.333 \\ 
  & Default+GAM     & $-$0.000 & 0.080 & 0.082 & 0.954 & 2.037 && 0.001 & 0.072 & 0.071 & 0.950 & 2.306 \\ 
  & Optimal         & $-$0.001 & 0.079 & 0.081 & 0.944 & 2.098 && $-$0.000 & 0.071 & 0.070 & 0.948 & 2.362 \\ \hline
	\end{tabular}}
    \begin{tabnote}
    NOTE: See the note to Table~\ref{tab1}.
    \end{tabnote}
	\label{stab1}
\end{table}

\begin{table}[htbp]\small
\tbl{Simulation results with multiple covariates under the Cox model (MCAR)}
	{
        \renewcommand{\tabcolsep}{4bp}
    \renewcommand{\arraystretch}{1.2}
\begin{tabular}{clrcc cccrc ccc}
		\hline
		&                 &     \multicolumn{5}{c}{$\beta_1=\log(2)$}      &  &    \multicolumn{5}{c}{$\beta_2=0.5$}    \\ \cline{3-7}\cline{9-13}
		$\rho$ & Method          &     Bias &  SSD  &  ESE  & CP   &  RE  &  &     Bias &  SSD  & ESE   &  CP  &  RE  \\ \hline
  &                 &        \multicolumn{11}{c}{Covariate Setting I} \\
  0.3   & Original        & 0.014 & 0.119 & 0.120 & 0.950 & 1.000 && 0.007 & 0.088 & 0.097 & 0.958 & 1.000 \\ 
  & Default         & 0.016 & 0.119 & 0.120 & 0.958 & 1.000 && $-$0.000 & 0.068 & 0.075 & 0.954 & 1.690 \\ 
  & Linear          & 0.029 & 0.122 & 0.127 & 0.952 & 0.959 && $-$0.020 & 0.076 & 0.087 & 0.974 & 1.373 \\ 
  & Default+Linear  & 0.025 & 0.123 & 0.132 & 0.962 & 0.932 && $-$0.007 & 0.071 & 0.085 & 0.974 & 1.552 \\ 
  & GAM             & 0.022 & 0.120 & 0.122 & 0.950 & 0.990 && $-$0.016 & 0.077 & 0.083 & 0.956 & 1.315 \\ 
  & Default+GAM     & 0.018 & 0.122 & 0.130 & 0.970 & 0.956 && $-$0.004 & 0.073 & 0.082 & 0.966 & 1.487 \\ 
  & Optimal         & 0.026 & 0.119 & 0.125 & 0.954 & 0.999 && $-$0.015 & 0.075 & 0.087 & 0.978 & 1.387 \\ 
  0.7   & Original        & 0.014 & 0.119 & 0.120 & 0.950 & 1.000 && 0.007 & 0.088 & 0.097 & 0.958 & 1.000 \\ 
  & Default         & 0.017 & 0.114 & 0.115 & 0.966 & 1.098 && $-$0.001 & 0.067 & 0.073 & 0.952 & 1.768 \\ 
  & Linear          & 0.026 & 0.112 & 0.116 & 0.958 & 1.139 && $-$0.011 & 0.067 & 0.074 & 0.960 & 1.759 \\ 
  & Default+Linear  & 0.022 & 0.113 & 0.119 & 0.962 & 1.115 && $-$0.002 & 0.067 & 0.076 & 0.966 & 1.723 \\ 
  & GAM             & 0.016 & 0.110 & 0.114 & 0.964 & 1.162 && $-$0.007 & 0.067 & 0.073 & 0.964 & 1.767 \\ 
  & Default+GAM     & 0.013 & 0.111 & 0.115 & 0.966 & 1.152 && 0.001 & 0.067 & 0.075 & 0.966 & 1.746 \\ 
  & Optimal         & 0.023 & 0.111 & 0.116 & 0.968 & 1.153 && $-$0.010 & 0.066 & 0.074 & 0.966 & 1.784 \\ \hline
  &                 &        \multicolumn{11}{c}{Covariate Setting II} \\
  0.3   & Original        & 0.008 & 0.099 & 0.100 & 0.946 & 1.000 && 0.012 & 0.070 & 0.068 & 0.936 & 1.000 \\ 
  & Default         & 0.006 & 0.099 & 0.100 & 0.952 & 1.009 && 0.006 & 0.056 & 0.055 & 0.936 & 1.548 \\ 
  & Linear          & 0.016 & 0.097 & 0.103 & 0.958 & 1.061 && 0.001 & 0.055 & 0.059 & 0.960 & 1.595 \\ 
  & Default+Linear  & 0.010 & 0.097 & 0.105 & 0.956 & 1.048 && 0.007 & 0.056 & 0.059 & 0.950 & 1.576 \\ 
  & GAM             & 0.003 & 0.079 & 0.082 & 0.954 & 1.574 && 0.003 & 0.047 & 0.049 & 0.958 & 2.195 \\ 
  & Default+GAM     & $-$0.004 & 0.080 & 0.083 & 0.950 & 1.541 && 0.010 & 0.049 & 0.051 & 0.946 & 2.067 \\ 
  & Optimal         & 0.005 & 0.077 & 0.082 & 0.956 & 1.673 && 0.003 & 0.046 & 0.048 & 0.956 & 2.282 \\ 
  0.7   & Original        & 0.008 & 0.099 & 0.100 & 0.946 & 1.000 && 0.012 & 0.070 & 0.068 & 0.936 & 1.000 \\ 
  & Default         & 0.007 & 0.095 & 0.096 & 0.950 & 1.096 && 0.007 & 0.054 & 0.054 & 0.942 & 1.700 \\ 
  & Linear          & 0.014 & 0.093 & 0.097 & 0.956 & 1.138 && 0.001 & 0.053 & 0.056 & 0.958 & 1.717 \\ 
  & Default+Linear  & 0.010 & 0.094 & 0.097 & 0.958 & 1.130 && 0.006 & 0.053 & 0.056 & 0.952 & 1.741 \\ 
  & GAM             & 0.001 & 0.078 & 0.080 & 0.952 & 1.616 && 0.004 & 0.047 & 0.048 & 0.954 & 2.257 \\ 
  & Default+GAM     & $-$0.003 & 0.079 & 0.081 & 0.956 & 1.583 && 0.009 & 0.048 & 0.050 & 0.952 & 2.159 \\ 
  & Optimal         & 0.003 & 0.076 & 0.080 & 0.954 & 1.717 && 0.004 & 0.046 & 0.048 & 0.956 & 2.328 \\ \hline
	\end{tabular}}
    \begin{tabnote}
    NOTE: See the note to Table~\ref{tab1}.
    \end{tabnote}
	\label{stab2}
\end{table}

\FloatBarrier

In the third simulation study, we consider two-phase sampling under MAR, that is, the missingness of $X$ depends on $Y$. All settings remain the same as in the first and second simulation studies except for the sampling schemes. Recall that the full sample size is $n=1000$ and the subsample with complete data has a size of $400$. Here, the subsample is selected depending on the outcome. For the continuous outcome $Y$, we divide subjects into two strata: those with $Y$ values above the upper $30$th percentile and those with $Y$ values below it. We sample $280$ subjects from the first stratum and sample $120$ subjects from the second stratum. For the binary outcome $Y$, we sample $280$ subjects from the stratum where $Y=1$ and sample $120$ subjects from the stratum where $Y=0$. For the survival outcome $Y=(T,\Delta)$, we sample $280$ subjects from the stratum where $\Delta=1$ (i.e., cases) and $120$ subjects from the stratum where $\Delta=0$ (i.e., controls). The sampling weights are calculated accordingly and used in our estimation procedure. The results are presented in Tables~\ref{stab3}--\ref{stab6}, corresponding to the settings of Tables~\ref{tab1}--\ref{stab2}, respectively. The conclusions remain the same as in Tables~\ref{tab1}--\ref{stab2}.

\begin{table}[htbp]\small
\tbl{Simulation results with one covariate (MAR)}
	{
    \renewcommand{\tabcolsep}{2bp}
    \renewcommand{\arraystretch}{1.2}
    \begin{tabular}{clrcc cccrc ccccr cccc}
		\hline
		&    &    \multicolumn{5}{c}{Linear Model}  & &   \multicolumn{5}{c}{Logistic Model}    &  &     \multicolumn{5}{c}{Cox Model}    \\ \cline{3-7}\cline{9-13}\cline{15-19}
		$\rho$ & Method   &     Bias &  SSD  &  ESE  & CP   &  RE  &  &     Bias &  SSD  & ESE   &  CP  &  RE  &  &     Bias &  SSD  & ESE   &  CP  &  RE \\ \hline
        &                 &        \multicolumn{17}{c}{Continuous Covariate} \\
		0.3   & Original &   0.001 &0.033 &0.031 &0.930 &1.000 & & $-$0.002 &0.111 &0.120 &0.968 &1.000 &  &  0.009 &0.077 &0.078 &0.936 &1.000  \\
		& Default  &   $-$0.001 &0.032 &0.031 &0.926 &1.049  & & $-$0.002 &0.108 &0.116 &0.964 &1.068 &  &  0.007 &0.077 &0.078 &0.940 &1.005  \\
		& Optimal  & 0.000 &0.027 &0.026 &0.932 &1.447 & & $-$0.011 &0.107 &0.115 &0.964 &1.083 &  & 0.010 &0.078 &0.079 &0.948 &0.992  \\
		0.7   & Original &  0.001 &0.033 &0.031 &0.930 &1.000 & & $-$0.002 &0.111 &0.119 &0.966 &1.000 &  & 0.009 &0.077 &0.078 &0.936 &1.000  \\
		& Default  &  0.000 &0.028 &0.028 &0.948 &1.346 & & $-$0.004 &0.093 &0.100 &0.956 &1.423 &  & 0.007 &0.071 &0.072 &0.936 &1.189  \\
		& Optimal  & 0.001 &0.026 &0.025 &0.934 &1.566 & & $-$0.009 &0.092 &0.099 &0.958 &1.448 &  &  0.010 &0.071 &0.071 &0.942 &1.202 \\ \hline
        &                 &        \multicolumn{17}{c}{Discrete Covariate} \\
		0.1   & Original &   0.000 &0.039 &0.038 &0.956 &1.000 & & 0.002 &0.129 &0.135 &0.958 &1.000 &  &  0.010 &0.176 &0.163 &0.926 &1.000  \\
		& Default  &   $-$0.001 &0.032 &0.031 &0.952 &1.461  & & $-$0.002 &0.109 &0.110 &0.954 &1.406 &  &  0.003 &0.130 &0.125 &0.934 &1.833  \\
		& Optimal  & 0.000 &0.031 &0.030 &0.948 &1.550 & & $-$0.004 &0.108 &0.109 &0.956 &1.434 &  & 0.002 &0.130 &0.127 &0.938 &1.822  \\
		0.3   & Original &   0.000 &0.039 &0.038 &0.956 &1.000 & & 0.002 &0.129 &0.135 &0.956 &1.000 &  & 0.010 &0.176 &0.163 &0.926 &1.000  \\
		& Default  &  $-$0.001 &0.038 &0.037 &0.950 &1.047 & & $-$0.003 &0.130 &0.132 &0.938 &0.980 &  & 0.001 &0.168 &0.155 &0.928 &1.091  \\
		& Optimal  & 0.000 &0.036 &0.036 &0.950 &1.140 & & $-$0.004 &0.126 &0.128 &0.940 &1.053 &  &  0.007 &0.169 &0.168 &0.952 &1.082 \\ \hline
	\end{tabular}}
    \begin{tabnote}
    NOTE: See the note to Table~\ref{tab1}.
    \end{tabnote}
	\label{stab3}
\end{table}

\begin{table}[htbp]\small
\tbl{Simulation results with multiple covariates under the linear model (MAR)}
	{
        \renewcommand{\tabcolsep}{4bp}
    \renewcommand{\arraystretch}{1.2}
    \begin{tabular}{clrcc cccrc ccc}
		\hline
		&                 &     \multicolumn{5}{c}{$\beta_1=1$}      &  &    \multicolumn{5}{c}{$\beta_2=0.5$}    \\ \cline{3-7}\cline{9-13}
		$\rho$ & Method          &     Bias &  SSD  &  ESE  & CP   &  RE  &  &     Bias &  SSD  & ESE   &  CP  &  RE  \\ \hline
  &                 &        \multicolumn{11}{c}{Covariate Setting I} \\
  0.3   & Original        & 0.002 & 0.048 & 0.050 & 0.944 & 1.000 && $-$0.000 & 0.041 & 0.041 & 0.928 & 1.000 \\ 
  & Default         & $-$0.002 & 0.048 & 0.049 & 0.942 & 0.993 && 0.002 & 0.037 & 0.037 & 0.932 & 1.196 \\ 
  & Linear          & 0.012 & 0.041 & 0.041 & 0.934 & 1.361 && $-$0.006 & 0.035 & 0.034 & 0.928 & 1.367 \\ 
  & Default+Linear  & 0.011 & 0.041 & 0.042 & 0.942 & 1.366 && $-$0.005 & 0.035 & 0.035 & 0.934 & 1.364 \\ 
  & GAM             & 0.015 & 0.042 & 0.042 & 0.936 & 1.333 && $-$0.009 & 0.035 & 0.035 & 0.926 & 1.338 \\ 
  & Default+GAM     & 0.014 & 0.042 & 0.042 & 0.932 & 1.346 && $-$0.007 & 0.035 & 0.035 & 0.932 & 1.356 \\ 
  & Optimal         & 0.012 & 0.041 & 0.041 & 0.934 & 1.373 && $-$0.006 & 0.035 & 0.034 & 0.928 & 1.370 \\ 
  0.7   & Original        & 0.002 & 0.048 & 0.050 & 0.944 & 1.000 && $-$0.000 & 0.041 & 0.041 & 0.928 & 1.000 \\ 
  & Default         & $-$0.001 & 0.047 & 0.048 & 0.938 & 1.047 && 0.001 & 0.036 & 0.036 & 0.946 & 1.266 \\ 
  & Linear          & 0.011 & 0.041 & 0.041 & 0.940 & 1.382 && $-$0.006 & 0.034 & 0.034 & 0.932 & 1.433 \\ 
  & Default+Linear  & 0.010 & 0.041 & 0.042 & 0.942 & 1.382 && $-$0.005 & 0.034 & 0.034 & 0.938 & 1.433 \\ 
  & GAM             & 0.012 & 0.042 & 0.042 & 0.942 & 1.348 && $-$0.007 & 0.035 & 0.034 & 0.934 & 1.408 \\ 
  & Default+GAM     & 0.011 & 0.041 & 0.042 & 0.944 & 1.351 && $-$0.006 & 0.034 & 0.034 & 0.930 & 1.423 \\ 
  & Optimal         & 0.011 & 0.041 & 0.041 & 0.938 & 1.392 && $-$0.006 & 0.034 & 0.034 & 0.934 & 1.437 \\ \hline
  &                 &        \multicolumn{11}{c}{Covariate Setting II} \\
  0.3   & Original        & 0.001 & 0.021 & 0.021 & 0.962 & 1.000 && 0.001 & 0.017 & 0.017 & 0.940 & 1.000 \\ 
  & Default         & 0.000 & 0.021 & 0.021 & 0.954 & 0.992 && 0.001 & 0.017 & 0.016 & 0.952 & 1.068 \\ 
  & Linear          & 0.003 & 0.020 & 0.020 & 0.936 & 1.095 && 0.002 & 0.016 & 0.016 & 0.940 & 1.174 \\ 
  & Default+Linear  & 0.003 & 0.020 & 0.020 & 0.936 & 1.085 && 0.002 & 0.016 & 0.016 & 0.954 & 1.169 \\ 
  & GAM             & 0.003 & 0.018 & 0.018 & 0.952 & 1.337 && 0.001 & 0.015 & 0.015 & 0.948 & 1.390 \\ 
  & Default+GAM     & 0.003 & 0.018 & 0.018 & 0.942 & 1.307 && 0.001 & 0.015 & 0.015 & 0.956 & 1.383 \\ 
  & Optimal         & 0.000 & 0.016 & 0.017 & 0.948 & 1.633 && 0.002 & 0.014 & 0.014 & 0.942 & 1.486 \\ 
  0.7   & Original        & 0.001 & 0.021 & 0.021 & 0.962 & 1.000 && 0.001 & 0.017 & 0.017 & 0.940 & 1.000 \\ 
  & Default         & 0.000 & 0.020 & 0.021 & 0.946 & 1.029 && 0.001 & 0.016 & 0.016 & 0.948 & 1.111 \\ 
  & Linear          & 0.003 & 0.020 & 0.020 & 0.930 & 1.083 && 0.002 & 0.016 & 0.016 & 0.948 & 1.208 \\ 
  & Default+Linear  & 0.003 & 0.020 & 0.020 & 0.940 & 1.097 && 0.002 & 0.016 & 0.016 & 0.950 & 1.211 \\ 
  & GAM             & 0.003 & 0.018 & 0.018 & 0.942 & 1.305 && 0.001 & 0.015 & 0.015 & 0.950 & 1.418 \\ 
  & Default+GAM     & 0.003 & 0.018 & 0.018 & 0.940 & 1.310 && 0.001 & 0.015 & 0.015 & 0.942 & 1.415 \\ 
  & Optimal         & 0.001 & 0.016 & 0.017 & 0.952 & 1.647 && 0.002 & 0.014 & 0.014 & 0.950 & 1.510 \\ \hline
	\end{tabular}}
    \begin{tabnote}
    NOTE: See the note to Table~\ref{tab1}.
    \end{tabnote}
	\label{stab4}
\end{table}

\begin{table}[htbp]\small
\tbl{Simulation results with multiple covariates under the logistic model (MAR)}
	{
        \renewcommand{\tabcolsep}{4bp}
    \renewcommand{\arraystretch}{1.2}
    \begin{tabular}{clrcc cccrc ccc}
		\hline
		&                 &     \multicolumn{5}{c}{$\beta_1=1$}      &  &    \multicolumn{5}{c}{$\beta_2=0.5$}    \\ \cline{3-7}\cline{9-13}
		$\rho$ & Method          &     Bias &  SSD  &  ESE  & CP   &  RE  &  &     Bias &  SSD  & ESE   &  CP  &  RE  \\ \hline
  &                 &        \multicolumn{11}{c}{Covariate Setting I} \\
  0.3   & Original        & 0.018 & 0.208 & 0.222 & 0.960 & 1.000 && 0.020 & 0.178 & 0.174 & 0.936 & 1.000 \\ 
  & Default         & 0.018 & 0.208 & 0.221 & 0.962 & 1.005 && $-$0.000 & 0.120 & 0.124 & 0.956 & 2.184 \\ 
  & Linear          & 0.016 & 0.210 & 0.222 & 0.960 & 0.990 && $-$0.016 & 0.119 & 0.128 & 0.954 & 2.222 \\ 
  & Default+Linear  & 0.016 & 0.210 & 0.232 & 0.970 & 0.989 && 0.001 & 0.122 & 0.129 & 0.958 & 2.114 \\ 
  & GAM             & $-$0.011 & 0.207 & 0.215 & 0.966 & 1.012 && $-$0.014 & 0.119 & 0.127 & 0.948 & 2.225 \\ 
  & Default+GAM     & $-$0.011 & 0.208 & 0.223 & 0.970 & 1.000 && 0.010 & 0.120 & 0.128 & 0.960 & 2.185 \\ 
  & Optimal         & 0.017 & 0.209 & 0.221 & 0.960 & 0.995 && $-$0.018 & 0.121 & 0.128 & 0.950 & 2.181 \\ 
  0.7   & Original        & 0.018 & 0.208 & 0.221 & 0.960 & 1.000 && 0.020 & 0.178 & 0.174 & 0.942 & 1.000 \\ 
  & Default         & 0.012 & 0.195 & 0.209 & 0.962 & 1.140 && 0.002 & 0.117 & 0.119 & 0.960 & 2.306 \\ 
  & Linear          & 0.001 & 0.195 & 0.210 & 0.966 & 1.148 && $-$0.003 & 0.115 & 0.119 & 0.958 & 2.378 \\ 
  & Default+Linear  & 0.003 & 0.196 & 0.216 & 0.970 & 1.134 && 0.003 & 0.117 & 0.124 & 0.964 & 2.318 \\ 
  & GAM             & $-$0.022 & 0.193 & 0.202 & 0.956 & 1.164 && 0.004 & 0.116 & 0.118 & 0.952 & 2.360 \\ 
  & Default+GAM     & $-$0.030 & 0.193 & 0.205 & 0.964 & 1.172 && 0.018 & 0.117 & 0.121 & 0.960 & 2.329 \\ 
  & Optimal         & 0.001 & 0.195 & 0.209 & 0.966 & 1.146 && $-$0.004 & 0.116 & 0.119 & 0.958 & 2.360 \\ \hline
  &                 &        \multicolumn{11}{c}{Covariate Setting II} \\
  0.3   & Original        & 0.034 & 0.124 & 0.123 & 0.948 & 1.000 && 0.024 & 0.124 & 0.131 & 0.952 & 1.000 \\ 
  & Default         & 0.033 & 0.124 & 0.124 & 0.956 & 1.011 && 0.011 & 0.089 & 0.095 & 0.962 & 1.932 \\ 
  & Linear          & $-$0.010 & 0.103 & 0.106 & 0.950 & 1.461 && $-$0.006 & 0.081 & 0.087 & 0.970 & 2.355 \\ 
  & Default+Linear  & $-$0.007 & 0.102 & 0.106 & 0.958 & 1.491 && $-$0.002 & 0.081 & 0.088 & 0.966 & 2.316 \\ 
  & GAM             & $-$0.011 & 0.081 & 0.086 & 0.958 & 2.335 && $-$0.011 & 0.075 & 0.080 & 0.960 & 2.724 \\ 
  & Default+GAM     & $-$0.008 & 0.082 & 0.087 & 0.960 & 2.313 && $-$0.009 & 0.075 & 0.081 & 0.960 & 2.721 \\ 
  & Optimal         & $-$0.006 & 0.082 & 0.086 & 0.944 & 2.307 && $-$0.009 & 0.074 & 0.079 & 0.960 & 2.783 \\ 
  0.7   & Original        & 0.034 & 0.124 & 0.123 & 0.948 & 1.000 && 0.024 & 0.124 & 0.131 & 0.952 & 1.000 \\ 
  & Default         & 0.024 & 0.113 & 0.112 & 0.942 & 1.203 && 0.008 & 0.085 & 0.090 & 0.966 & 2.113 \\ 
  & Linear          & $-$0.011 & 0.097 & 0.100 & 0.948 & 1.631 && $-$0.007 & 0.079 & 0.085 & 0.966 & 2.441 \\ 
  & Default+Linear  & $-$0.004 & 0.096 & 0.100 & 0.952 & 1.677 && $-$0.000 & 0.080 & 0.087 & 0.970 & 2.414 \\ 
  & GAM             & $-$0.010 & 0.081 & 0.085 & 0.954 & 2.387 && $-$0.011 & 0.074 & 0.080 & 0.964 & 2.773 \\ 
  & Default+GAM     & $-$0.006 & 0.081 & 0.086 & 0.956 & 2.336 && $-$0.007 & 0.074 & 0.081 & 0.968 & 2.760 \\ 
  & Optimal         & $-$0.007 & 0.080 & 0.085 & 0.954 & 2.403 && $-$0.009 & 0.074 & 0.079 & 0.962 & 2.826 \\ \hline
	\end{tabular}}
    \begin{tabnote}
    NOTE: See the note to Table~\ref{tab1}.
    \end{tabnote}
	\label{stab5}
\end{table}

\begin{table}[htbp]\small
\tbl{Simulation results with multiple covariates under the Cox model (MAR)}
	{
        \renewcommand{\tabcolsep}{4bp}
    \renewcommand{\arraystretch}{1.2}
    \begin{tabular}{clrcc cccrc ccc}
		\hline
		&                 &     \multicolumn{5}{c}{$\beta_1=\log(2)$}      &  &    \multicolumn{5}{c}{$\beta_2=0.5$}    \\ \cline{3-7}\cline{9-13}
		$\rho$ & Method          &     Bias &  SSD  &  ESE  & CP   &  RE  &  &     Bias &  SSD  & ESE   &  CP  &  RE  \\ \hline
  &                 &        \multicolumn{11}{c}{Covariate Setting I} \\
  0.3   & Original        & 0.015 & 0.119 & 0.118 & 0.938 & 1.000 && 0.004 & 0.095 & 0.098 & 0.956 & 1.000 \\ 
  & Default         & 0.008 & 0.119 & 0.117 & 0.932 & 1.009 && 0.004 & 0.074 & 0.077 & 0.958 & 1.654 \\ 
  & Linear          & 0.031 & 0.123 & 0.127 & 0.948 & 0.941 && $-$0.016 & 0.078 & 0.086 & 0.956 & 1.473 \\ 
  & Default+Linear  & 0.028 & 0.124 & 0.131 & 0.946 & 0.925 && $-$0.004 & 0.076 & 0.086 & 0.966 & 1.588 \\ 
  & GAM             & 0.021 & 0.122 & 0.121 & 0.936 & 0.962 && $-$0.010 & 0.077 & 0.083 & 0.952 & 1.520 \\ 
  & Default+GAM     & 0.019 & 0.121 & 0.126 & 0.948 & 0.966 && $-$0.002 & 0.074 & 0.082 & 0.966 & 1.660 \\ 
  & Optimal         & 0.025 & 0.121 & 0.124 & 0.958 & 0.978 && $-$0.010 & 0.076 & 0.086 & 0.972 & 1.568 \\ 
  0.7   & Original        & 0.015 & 0.119 & 0.118 & 0.938 & 1.000 && 0.004 & 0.095 & 0.098 & 0.956 & 1.000 \\ 
  & Default         & 0.009 & 0.114 & 0.112 & 0.934 & 1.103 && 0.004 & 0.073 & 0.075 & 0.962 & 1.724 \\ 
  & Linear          & 0.027 & 0.114 & 0.115 & 0.938 & 1.102 && $-$0.009 & 0.072 & 0.076 & 0.956 & 1.744 \\ 
  & Default+Linear  & 0.024 & 0.114 & 0.118 & 0.950 & 1.091 && 0.001 & 0.073 & 0.078 & 0.962 & 1.690 \\ 
  & GAM             & 0.017 & 0.112 & 0.112 & 0.938 & 1.129 && $-$0.005 & 0.071 & 0.074 & 0.948 & 1.800 \\ 
  & Default+GAM     & 0.015 & 0.113 & 0.113 & 0.946 & 1.114 && 0.003 & 0.072 & 0.076 & 0.958 & 1.756 \\ 
  & Optimal         & 0.023 & 0.112 & 0.114 & 0.938 & 1.127 && $-$0.007 & 0.071 & 0.075 & 0.956 & 1.804 \\ \hline
  &                 &        \multicolumn{11}{c}{Covariate Setting II} \\
  0.3   & Original        & 0.002 & 0.094 & 0.094 & 0.944 & 1.000 && 0.014 & 0.063 & 0.062 & 0.934 & 1.000 \\ 
  & Default         & $-$0.007 & 0.092 & 0.093 & 0.954 & 1.043 && 0.013 & 0.054 & 0.053 & 0.936 & 1.360 \\ 
  & Linear          & 0.007 & 0.093 & 0.097 & 0.958 & 1.026 && 0.009 & 0.054 & 0.056 & 0.944 & 1.357 \\ 
  & Default+Linear  & 0.000 & 0.092 & 0.098 & 0.960 & 1.038 && 0.015 & 0.055 & 0.056 & 0.940 & 1.311 \\ 
  & GAM             & 0.003 & 0.078 & 0.079 & 0.942 & 1.451 && 0.009 & 0.048 & 0.048 & 0.952 & 1.728 \\ 
  & Default+GAM     & $-$0.005 & 0.078 & 0.079 & 0.956 & 1.444 && 0.015 & 0.049 & 0.049 & 0.930 & 1.673 \\ 
  & Optimal         & 0.004 & 0.076 & 0.077 & 0.950 & 1.540 && 0.009 & 0.047 & 0.047 & 0.944 & 1.788 \\ 
  0.7   & Original        & 0.002 & 0.094 & 0.094 & 0.944 & 1.000 && 0.014 & 0.063 & 0.062 & 0.934 & 1.000 \\ 
  & Default         & $-$0.003 & 0.088 & 0.089 & 0.954 & 1.137 && 0.013 & 0.052 & 0.052 & 0.938 & 1.469 \\ 
  & Linear          & 0.009 & 0.089 & 0.092 & 0.964 & 1.121 && 0.008 & 0.052 & 0.053 & 0.954 & 1.446 \\ 
  & Default+Linear  & 0.004 & 0.087 & 0.092 & 0.964 & 1.150 && 0.013 & 0.053 & 0.053 & 0.944 & 1.429 \\ 
  & GAM             & 0.004 & 0.076 & 0.078 & 0.954 & 1.532 && 0.009 & 0.047 & 0.047 & 0.950 & 1.767 \\ 
  & Default+GAM     & $-$0.002 & 0.076 & 0.078 & 0.952 & 1.526 && 0.014 & 0.048 & 0.048 & 0.934 & 1.742 \\ 
  & Optimal         & 0.004 & 0.074 & 0.076 & 0.954 & 1.609 && 0.009 & 0.047 & 0.047 & 0.944 & 1.824 \\ \hline
	\end{tabular}}
    \begin{tabnote}
    NOTE: See the note to Table~\ref{tab1}.
    \end{tabnote}
	\label{stab6}
\end{table}

\FloatBarrier

We conduct some additional simulation studies to further assess the proposed method. First, we evaluate the choice of bandwidth for kernel estimation in the optimal update method under the linear model with a continuous covariate as in the first simulation study. We consider three values of bandwidth: $a_n=\sigma_{X^*} \, n^{-1/6}$, $1.06 \sigma_{X^*} \, n^{-1/5}$, and $\sigma_{X^*} \, n^{-1/4}$. The results are presented in Table~\ref{stab7}. One can see that the proposed method yields very similar results under different values of bandwidth.

\begin{table}[htbp]\small
\tbl{Simulation results with different bandwidths under the linear model}
	{
    \renewcommand{\tabcolsep}{2bp}
    \renewcommand{\arraystretch}{1.2}
    \begin{tabular}{clrcc cccrc ccccr cccc}
		\hline
		&    &    \multicolumn{5}{c}{$a_n=\sigma_{X^*} \, n^{-1/6}$}  & &   \multicolumn{5}{c}{$a_n=1.06 \, \sigma_{X^*} \, n^{-1/5}$}    &  &     \multicolumn{5}{c}{$a_n=\sigma_{X^*} \, n^{-1/4}$}    \\ \cline{3-7}\cline{9-13}\cline{15-19}
		$\rho$ & Method   &     Bias &  SSD  &  ESE  & CP   &  RE  &  &     Bias &  SSD  & ESE   &  CP  &  RE  &  &     Bias &  SSD  & ESE   &  CP  &  RE \\ \hline
		0.3   & Original & $-$0.001 &0.024 &0.025 &0.962 &1.000 && $-$0.001 &0.024 &0.025 &0.962 &1.000 & & $-$0.001 &0.024 &0.025 &0.962 &1.000   \\
		& Default  &  $-$0.001 &0.024 &0.025 &0.944 &1.041 && $-$0.001 &0.024 &0.025 &0.944 &1.041  & & $-$0.001 &0.024 &0.025 &0.944 &1.041  \\
		& Optimal  & $-$0.001 &0.021 &0.021 &0.952 &1.380 && $-$0.001 &0.021 &0.021 &0.952 &1.377 & & $-$0.002 &0.021 &0.021 &0.946 &1.370  \\
		0.7   & Original & $-$0.001 &0.024 &0.025 &0.962 &1.000 && $-$0.001 &0.024 &0.025 &0.962 &1.000 & & $-$0.001 &0.024 &0.025 &0.962 &1.000   \\
		& Default  & 0.000 &0.022 &0.022 &0.948 &1.268 && 0.000 &0.022 &0.022 &0.948 &1.268 & & 0.000 &0.022 &0.022 &0.948 &1.268   \\
		& Optimal  &  0.000 &0.020 &0.021 &0.948 &1.479 && 0.000 &0.020 &0.021 &0.952 &1.485 & & $-$0.001 &0.020 &0.021 &0.952 &1.492 \\ \hline
	\end{tabular}}
    \begin{tabnote}
    NOTE: See the note to Table~\ref{tab1}.
    \end{tabnote}
	\label{stab7}
\end{table}

Second, we compare the proposed optimal update estimator with the estimator of \cite{zeng2014efficient} under the Cox model with a discrete covariate as in the first simulation study. The results are given in Table~\ref{stab8}. As expected, the estimator of \cite{zeng2014efficient} is more efficient than the proposed estimator, since it is semiparametrically efficient whereas the proposed estimator is optimal only within the class of update estimators. However, the method of \cite{zeng2014efficient} is based on kernel estimation of $F_{X|Z}$ and suffers from the curse of dimensionality when the dimension of $Z$ is higher. In contrast, when the dimension of $Z$ is high, the update method can still be applied by specifying a working model for $F_{X|Z}$. We can also specify multiple plausible working models for $F_{X|Z}$ and perform joint update to enhance the estimation efficiency.

\begin{table}[htbp]\small
\tbl{Simulation results for comparison with Zeng--Lin}
	{
    \renewcommand{\tabcolsep}{2bp}
    \renewcommand{\arraystretch}{1.2}
    \begin{tabular}{clrcc cccrc ccc}
		\hline
		&    &   \multicolumn{5}{c}{MCAR}    &    &   \multicolumn{5}{c}{MAR}    \\ \cline{3-7}\cline{9-13}
		$\rho$ & Method   &     Bias &  SSD  &  ESE  & CP   &  RE  &   &  Bias &  SSD  &  ESE  & CP   &  RE  \\ \hline
		0.1   & Original &    0.010 &0.146 &0.148 &0.944 &1.000 &   & 0.010 &0.176 &0.163 &0.926 &1.000 \\
		       & Proposed  &   0.004 &0.118 &0.118 &0.958 &1.532 &   & 0.002 &0.130 &0.127 &0.938 &1.822 \\
		       & Zeng--Lin  &  $-$0.002 &0.115 &0.111 &0.938 &1.612 &   & $-$0.001 &0.116 &0.113 &0.944 &2.302 \\ 
        0.3   & Original &   0.010 &0.146 &0.148 &0.944 &1.000 &   & 0.010 &0.176 &0.163 &0.926 &1.000 \\
		       & Proposed  &   0.011 &0.142 &0.150 &0.956 &1.061  &   & 0.007 &0.169 &0.168 &0.952 &1.082 \\
		       & Zeng--Lin  &  $-$0.000 &0.140 &0.135 &0.938 &1.088 &   & $-$0.004 &0.147 &0.140 &0.934 &1.434 \\ \hline
	\end{tabular}}
    \begin{tabnote}
    NOTE: See the note to Table~\ref{tab1}.
    \end{tabnote}
	\label{stab8}
\end{table}

\FloatBarrier

\section{Application to the TCGA Study}

We are interested in identifying genes associated with the prognosis of ovarian cancer patients. We consider a data set from The Cancer Genome Atlas (TCGA) \citep{cancer2011integrated}, which is publicly available at https://gdac.broadinstitute.org. In this study, most subjects have genomic data available, including mRNA expressions measured by three microarray platforms. Also, for a subset of subjects, mRNA expressions were measured by RNA sequencing, a more advanced technology than microarray profiling. In addition, demographic and clinical variables were measured, including age at diagnosis, tumor stage, tumor grade, and time to death since initial diagnosis.

We excluded subjects with missing clinical or demographic variables, as well as those with tumor stage I or tumor grade 1. The resulting sample size is $n=450$, and the censoring rate for the survival time is about $39\%$. A subset of $278$ subjects also have available RNA sequencing data. There is no appreciable difference in the survival time or covariates between subjects with or without RNA sequencing data, and thus we treat the missing data as MCAR. In the dataset, there are 20531 genes with RNA sequencing data. For the three microarray platforms, namely Agilent, Affymetrix HuEx, and Affymetrix U133A, there are about 12000--18000 gene expression measurements. To obtain a summary gene expression from the microarray platforms, we follow \cite{cancer2011integrated} and fit a factor model with a single latent factor for the three microarray measurements, separately for each gene. Then, we set the summary expression for each subject as the estimated conditional expected value of the latent factor given the observed microarray measurements. We keep the genes for which at least two of the three microarray measurements are correlated with the summary expression by more than $0.7$. The number of genes with summary expression  meeting this criterion and with available RNA sequencing data is 10990.

In the analysis, we consider each gene separately and fit the Cox model on the expression measured by RNA sequencing, adjusting for age, tumor stage, and tumor grade. The summary microarray expression is used as an auxiliary variable. We compare four methods: the original estimator based on the subsample, the default update estimator, the linear update estimator, and the joint update estimator that combines the default and linear update estimators. The update estimators are obtained using the fully bootstrap method described in Section 2.4 with $200$ bootstrap samples. Out of the 10990 genes, 7 are selected at a significance level $\alpha=0.05$ after the Bonferroni correction for multiple testing. In particular, POU3F2, RPS6KA2, and SNX17 are selected by the linear update method; DAP is selected by the joint update method; GNAS, RBMS1, and SLC12A9 are selected by both the linear update and joint update methods; no genes are selected by the original method or the default update method. The results for the 7 selected genes are presented in Table~\ref{tab3}. One can see that the joint update method yields the smallest standard errors for all 7 genes. In fact, 2 out of these 7 genes are among the top 5 significant genes under the original method. This suggests that these genes exhibit evidence of association with the survival time under the original method, but the signals are not strong enough to reach significance due to the lack of efficiency of the original method.

\begin{table}
\tbl{Analysis results of TCGA for 7 significant genes selected by the Bonferroni correction}
{
\renewcommand{\tabcolsep}{2bp}
  \begin{tabular}{ccccc ccccc ccccc c}
  \hline
  &  \multicolumn{3}{c}{DAP} & & \multicolumn{3}{c}{GNAS}  & & \multicolumn{3}{c}{POU3F2}  & & \multicolumn{3}{c}{RBMS1} \\ \cline{2-4}\cline{6-8}\cline{10-12}\cline{14-16}
Method &  Est & SE & p-value & & Est & SE & p-value & & Est & SE & p-value & & Est & SE & p-value \\ \hline
Original & $-$0.715 & 0.173 & 3.45E$-$05 &  & $-$0.345 & 0.142 & 1.51E$-$02 &  & 0.110 & 0.064 & 8.44E$-$02 &  & 0.606 & 0.169 & 3.28E$-$04 \\ 
Default  & $-$0.503 & 0.128 & 8.18E$-$05 &  & $-$0.318 & 0.112 & 4.29E$-$03 &  & 0.093 & 0.056 & 1.00E$-$01 &  & 0.381 & 0.147 & 9.87E$-$03 \\ 
Linear   & $-$0.752 & 0.170 & 1.01E$-$05 &  & $-$0.520 & 0.108 & 1.38E$-$06 &  & $-$0.319 & 0.056 & 1.38E$-$08 &  & 0.719 & 0.141 & 3.25E$-$07 \\ 
Default+Linear   & $-$0.548 & 0.106 & 2.06E$-$07 &  & $-$0.403 & 0.085 & 1.99E$-$06 &  & $-$0.238 & 0.056 & 2.09E$-$05 &  & 0.582 & 0.121 & 1.45E$-$06\\ \hline
  &  \multicolumn{3}{c}{RPS6KA2}  & & \multicolumn{3}{c}{SLC12A9}  & & \multicolumn{3}{c}{SNX17} \\ \cline{2-4}\cline{6-8}\cline{10-12}
Method &  Est & SE & p-value & & Est & SE & p-value & & Est & SE & p-value \\ \hline
Original & 0.298 & 0.095 & 1.77E$-$03 &  & 0.454 & 0.136 & 8.80E$-$04 &  & $-$0.676 & 0.201 & 7.64E$-$04 \\ 
Default  & 0.260 & 0.071 & 2.43E$-$04 &  & 0.304 & 0.111 & 6.15E$-$03 &  & $-$0.289 & 0.184 & 1.17E$-$01 \\ 
Linear   & 0.328 & 0.072 & 4.44E$-$06 &  & 0.516 & 0.097 & 1.15E$-$07 &  & $-$0.880 & 0.191 & 4.09E$-$06 \\ 
Default+Linear   & 0.299 & 0.066 & 6.15E$-$06 &  & 0.422 & 0.084 & 5.40E$-$07 &  & $-$0.517 & 0.140 & 2.22E$-$04\\\hline
  \end{tabular}}
\label{tab3}
\end{table}

\section{Discussion}

We propose a robust method to improve estimation efficiency in two-phase studies. The proposed optimal and joint update estimators are more efficient than the complete-data estimator and the standard update estimator, regardless of the correctness of the working models specified in the estimation procedure. The proposed methods are based on influence functions and are generally applicable. Although we focus on two-phase studies, the proposed methods can be applied to more general missing data problems. In fact, as long as one can find a complete-data estimator that is asymptotically linear, the proposed methods can be used to improve the estimation efficiency.

One interesting but challenging case is when data are missing not at random (MNAR). Although MNAR is not common in two-phase studies, it is frequently encountered in some other situations. In general, handling MNAR data is a highly challenging problem, because the missingness mechanism depends on the unobserved value itself and the observed data alone generally do not identify the full data distribution. One possible approach is to use an IPW initial estimator with a consistently estimated MNAR response probability. Under MNAR, this response probability depends on the missing variable itself and cannot be estimated by standard methods. Auxiliary variables and suitable structural assumptions on the missing mechanism, such as those considered in the instrumental-variable framework of \cite{sun2018semiparametric}, could be adopted to estimate the weights. Once such weights are consistently estimated, the resulting IPW estimator could serve as the initial estimator in the proposed update procedure. Relevant discussion on applying update estimation to MNAR settings can also be found in \cite{cannings2022correlation}.

We demonstrate the proposed methods using the linear model, logistic model, and Cox model with right-censored data. Similar approaches can be developed to improve update estimators for other outcome models and data structures, such as the additive hazards model, accelerated failure time model, and survival models for interval-censored data. In these cases, formulation of the influence function of the original estimator, and thus the optimal working update term, may be challenging. Nevertheless, even without developing an optimal update approach, existing methods can be improved by employing the joint update strategy.  Specifically, multiple plausible working models can be utilized to construct an update term. These models can belong to different classes or share the same structure but differ in the specification of nuisance parameters, such as whether a baseline hazard is modeled nonparametrically or parametrically.

The proposed joint update approach is reminiscent of the ``multiply robust'' estimator \citep{han2013estimation, han2014multiply}. This estimator extends the AIPW estimators by incorporating multiple propensity score models and outcome regression models, ensuring consistency if at least one of these models is correctly specified. In addition, if one propensity model and one outcome regression model are correctly specified, then the multiply robust estimator achieves optimal efficiency. The proposed joint update approach shares this optimality property: if one of the working estimators is optimal, then the proposed estimator attains optimal efficiency. Despite sharing these theoretical properties, the proposed approach is both conceptually and computationally simpler. Existing multiply robust estimators rely on empirical likelihood methods, which require solving constrained optimization problems. In contrast, the joint update estimator can be easily computed using existing methods if standard models are adopted for the original and working estimators.

There are several potential directions for future research. One is to extend the proposed methods to update infinite-dimensional parameters, such as the baseline hazard function in the Cox model. This extension would be of particular interest for event time prediction under the Cox model, as the survival probability depends on both the regression parameters and baseline hazard function. Another direction is to develop an update approach for high-dimensional regression for both inference and variable selection. Update estimation cannot be directly applied to popular penalized estimators, such as the lasso and elastic net estimators, because they are not asymptotically linear. To overcome this limitation, we can develop update estimation approaches based on asymptotically linear debiased estimators.

\section*{Acknowledgement}
Q. Zhou's work is partially supported by the National Science Foundation grant DMS1916170. K. Y. Wong's work is partially supported by the GuangDong Basic and Applied Basic Research Foundation (Project No. 2021A1515110048), a research grant from the Hong Kong Polytechnic University (1-ZVX4), and the Hong Kong Research Grants Council under Grant 15303422.

\section*{Software}
An R package, named \texttt{twoStep2Phase}, that implements the proposed methods is available at GitHub (alexwky/twoStep2Phase).

\clearpage
\appendix
\section*{Appendix}
\setcounter{equation}{0}
\renewcommand{\theequation}{A.\arabic{equation}}
\begin{proof}[Proof of Theorem 1]
Without loss of generality, assume that $\psi_0$ is one-dimensional. Note that
\[
\Sigma_{12}=\mathrm{E}\bigg\{\frac{R}{\pi}\psi_0\Big(\frac{R}{\pi}-1\Big)\phi_0^\mathrm{T}\bigg\}=\mathrm{E}\bigg\{\mathrm{Var}\Big(\frac{R}{\pi}\Big|Y,X,Z\Big)\psi_0\phi_0^{\mathrm{T}}\bigg\}=\mathrm{E}\bigg(\frac{1-\pi}{\pi}\psi_0\phi_0^{\mathrm{T}}\bigg)
\]
and
\[
\Sigma_{22}=\mathrm{E}\bigg\{\Big(\frac{R}{\pi}-1\Big)^2\phi_0\phi_0^\mathrm{T}\bigg\}=\mathrm{E}\bigg\{\mathrm{Var}\Big(\frac{R}{\pi}\Big|Y,X,Z\Big)\phi_0\phi_0^{\mathrm{T}}\bigg\}=\mathrm{E}\bigg(\frac{1-\pi}{\pi}\phi_0\phi_0^{\mathrm{T}}\bigg).
\]
Therefore,
\[
n^{1/2}(\overline{\theta}-\theta_0)=n^{1/2}\mathbb{P}_n\bigg\{\frac{R}{\pi}\psi_{0}-\mathrm{E}\bigg(\frac{1-\pi}{\pi}\psi_0\phi_0^{\mathrm{T}}\bigg)\mathrm{E}\bigg(\frac{1-\pi}{\pi}\phi_0\phi_0^{\mathrm{T}}\bigg)^{-1}\bigg(\frac{R}{\pi}-1\bigg)\phi_{0}\bigg\}+o_p(1).
\]
Clearly, the asymptotic distribution of $\overline{\theta}$ has the desired form.

To derive the optimal choice of $\phi_0$, note that
\[
\mathrm{E}\bigg(\frac{1-\pi}{\pi}\phi_0\psi_0\bigg)=\mathrm{E}\bigg\{\frac{1-\pi}{\pi}\phi_0\mathrm{E}(\psi_0\mid Y,Z)\bigg\}.
\]
The second term in $\Sigma(\phi_0)$ is equal to
\begin{align}
&\,\mathrm{E}\bigg\{\frac{1-\pi}{\pi}\mathrm{E}(\psi_0\mid Y,Z)\phi_0^{\mathrm{T}}\bigg\}\mathrm{E}\bigg(\frac{1-\pi}{\pi}\phi_0\phi_0^{\mathrm{T}}\bigg)^{-1} \mathrm{E}\bigg\{\frac{1-\pi}{\pi}\phi_0\mathrm{E}(\psi_0\mid Y,Z)\bigg\}\nonumber\\=&\, \Bigg\Vert\mathrm{E}\bigg\{\frac{1-\pi}{\pi}\mathrm{E}\bigg(\frac{1-\pi}{\pi}\phi_0\phi_0^{\mathrm{T}}\bigg)^{-1/2}\phi_0\mathrm{E}(\psi_0\mid Y,Z)\bigg\}\Bigg\Vert^2.\label{eq:varred}
\end{align}
First, we show that it suffices to consider only one-dimensional $\phi_0$'s. For any $\phi_0$ of dimension $p$, we can find a vector $a\in\mathbb{R}^p$ and a random variable $\varphi\equiv\varphi(Y,Z)$, such that
\[
\mathrm{E}(\psi_0\mid Y,Z)=a^{\mathrm{T}}\mathrm{E}\bigg(\frac{1-\pi}{\pi}\phi_0\phi_0^{\mathrm{T}}\bigg)^{-1/2}\phi_0+\varphi,
\]
and $\mathrm{E}\{(1-\pi)\pi^{-1}\varphi\phi_0\}=0$. Effectively, we are decomposing $\mathrm{E}(\psi_0\mid Y,Z)$ into a term from $\mathrm{span}\{\phi_{0j}:j=1,\ldots,p\}$ and a term from its orthogonal complement, where the inner product between two elements $\varphi_1$ and $\varphi_2$ is defined as $\mathrm{E}\{(1-\pi)\pi^{-1}\varphi_1\varphi_2\}$. For
\[
\overline{\phi}_0=a^{\mathrm{T}}\mathrm{E}\bigg(\frac{1-\pi}{\pi}\phi_0\phi_0^{\mathrm{T}}\bigg)^{-1/2}\phi_0,
\]
we have
\begin{align*}
\Bigg\Vert\mathrm{E}\bigg\{\frac{1-\pi}{\pi}\mathrm{E}\bigg(\frac{1-\pi}{\pi}\phi_0\phi_0^{\mathrm{T}}\bigg)^{-1/2}\phi_0\mathrm{E}(\psi_0\mid Y,Z)\bigg\}\Bigg\Vert^2
=\Bigg[\mathrm{E}\bigg\{\frac{1-\pi}{\pi}\mathrm{E}\bigg(\frac{1-\pi}{\pi}\overline{\phi}_0^2\bigg)^{-1/2}\overline{\phi}_0\mathrm{E}(\psi_0\mid Y,Z)\bigg\}\Bigg]^2,
\end{align*}
so any value of (\ref{eq:varred}) can be attained by a one-dimensional $\phi_0$.

Now, with a one-dimensional $\phi_0$, we can show by the Cauchy--Schwarz inequality that
\begin{align*}
\Bigg[\mathrm{E}\bigg\{\frac{1-\pi}{\pi}\mathrm{E}\bigg(\frac{1-\pi}{\pi}\phi_0^2\bigg)^{-1/2}\phi_0\mathrm{E}(\psi_0\mid Y,Z)\bigg\}\Bigg]^2
=&\,\Bigg[\mathrm{E}\bigg\{\bigg(\frac{R}{\pi}-1\bigg)^2\mathrm{E}\bigg(\frac{1-\pi}{\pi}\phi_0^2\bigg)^{-1/2}\phi_0\mathrm{E}(\psi_0\mid Y,Z)\bigg\}\Bigg]^2\\
\le&\,\mathrm{E}\bigg\{\frac{1-\pi}{\pi}\mathrm{E}(\psi_0\mid Y,Z)^2\bigg\}.
\end{align*}
The desired result follows from the fact that equality on the last line above is attained at $\phi_0=c\mathrm{E}(\psi_0\mid Y,Z)$ for any nonzero constant $c$.
\end{proof}

\begin{proof}[Proof of Theorem 2]
First, we show that $\widehat{\vartheta}_F\rightarrow_p\theta_0$ and $\widehat{\vartheta}_S\rightarrow_p\theta_0$. For the full-sample estimator, we have
\begin{align*}
0 & =\mathbb{P}_n\phi^*(Y,Z;\widehat{\vartheta}_F,\widehat{\eta},\widehat{F})\nonumber\\
 & =\mathbb{P}\phi^*(Y,Z;\widehat{\vartheta}_F,\eta_0,F_0)+(\mathbb{P}_n-\mathbb{P})\phi^*(Y,Z;\widehat{\vartheta}_F,\eta_0,F_0)
\nonumber\\
&\quad+\mathbb{P}_n\{\phi^*(Y,Z;\widehat{\vartheta}_F,\widehat{\eta},\widehat{F})-\phi^*(Y,Z;\widehat{\vartheta}_F,\eta_0,F_0)\}\\
 &\equiv \mathbb{P}\phi^*(Y,Z;\widehat{\vartheta}_F,\eta_0,F_0)+k_n.
\end{align*}
For any $\epsilon>0$, let $D_\epsilon=\{\theta\in\Theta:\Vert\theta-\theta_0\Vert>\epsilon\}$. By the continuity of $\phi^*$ in $\theta$ and the compactness of $\Theta$, we can find $\delta_\epsilon>0$ such that $\theta\in D_\epsilon\implies |\mathbb{P}\phi^*(Y,Z;\theta,\eta_0,F_0)|>\delta_\epsilon$. Therefore,
\[
P(\widehat{\vartheta}_F\in D_\epsilon)\le P(|\mathbb{P}\phi^*(Y,Z;\widehat{\vartheta}_F,\eta_0,F_0)|>\delta_\epsilon)=P(|k_n|>\delta_\epsilon),
\]
where the right-hand side tends to 0 by Condition 4. Therefore, $\widehat{\vartheta}_F$ is consistent. We can similarly show that $\widehat{\vartheta}_S$ is consistent.

Note that
\begin{align}
0 & =n^{1/2}\mathbb{P}_n\phi^*(Y,Z;\widehat{\vartheta}_F,\widehat{\eta},\widehat{F})\nonumber\\
 & =n^{1/2}\mathbb{P}_n\phi^*(Y,Z;\theta_0,\eta_0,F_0)+n^{1/2}(\widehat{\vartheta}_F-\theta_{0})^{\mathrm{T}}\mathbb{P}_n\phi^*_\theta(Y,Z;\theta_0,\eta_0,F_0)\nonumber\\
 &\quad+n^{1/2}\mathbb{P}_n\Big\{\phi^*(Y,Z;\theta_0,\widehat{\eta},\widehat{F})-\phi^*(Y,Z;\theta_0,\eta_0,F_0)\Big\}\nonumber\\
&\quad +n^{1/2}(\widehat{\vartheta}_F-\theta_0)^{\mathrm{T}}\mathbb{P}_n\Big\{\phi^*_\theta(Y,Z;\theta_0,\widehat{\eta},\widehat{F})-\phi^*_\theta(Y,Z;\theta_0,\eta_0,F_0)\Big\}\nonumber\\
&\quad+\frac{1}{2} n^{1/2}(\widehat{\vartheta}_F-\theta_0)^{\mathrm{T}}\mathbb{P}_n\phi^*_{\theta\theta}(Y,Z;\widetilde{\theta},\widehat{\eta},\widehat{F})(\widehat{\vartheta}_F-\theta_0)\nonumber\\
&=n^{1/2}\mathbb{P}_n\phi^*(Y,Z;\theta_0,\eta_0,F_0)+n^{1/2}(\widehat{\vartheta}_F-\theta_{0})^{\mathrm{T}}\bigg[\mathbb{P}\phi^*_\theta(Y,Z;\theta_0,\eta_0,F_0)+(\mathbb{P}_n-\mathbb{P})\phi^*_\theta(Y,Z;\theta_0,\eta_0,F_0)\nonumber\\
&\quad+\mathbb{P}\Big\{\phi^*_\theta(Y,Z;\theta_0,\widehat{\eta},\widehat{F})-\phi^*_\theta(Y,Z;\theta_0,\eta_0,F_0)\Big\}\nonumber\\&\quad+(\mathbb{P}_n-\mathbb{P})\Big\{\phi^*_\theta(Y,Z;\theta_0,\widehat{\eta},\widehat{F})-\phi^*_\theta(Y,Z;\theta_0,\eta_0,F_0)\Big\}\bigg]\nonumber\\
 &\quad+n^{1/2}\mathbb{P}\Big\{\phi^*(Y,Z;\theta_0,\widehat{\eta},\widehat{F})-\phi^*(Y,Z;\theta_0,\eta_0,F_0)\Big\}\nonumber\\
  &\quad+n^{1/2}(\mathbb{P}_n-\mathbb{P})\Big\{\phi^*(Y,Z;\theta_0,\widehat{\eta},\widehat{F})-\phi^*(Y,Z;\theta_0,\eta_0,F_0)\Big\}\nonumber\\
&\quad+\frac{1}{2} n^{1/2}(\widehat{\vartheta}_F-\theta_0)^{\mathrm{T}}\mathbb{P}_n\phi^*_{\theta\theta}(Y,Z;\widetilde{\theta},\widehat{\eta},\widehat{F})(\widehat{\vartheta}_F-\theta_0)\nonumber\\
&=n^{1/2}\mathbb{P}_n\phi^*(Y,Z;\theta_0,\eta_0,F_0)+n^{1/2}(\widehat{\vartheta}_F-\theta_{0})^{\mathrm{T}}\bigg\{\mathbb{P}\phi^*_\theta(Y,Z;\theta_0,\eta_0,F_0)+o_p(1)\bigg\}\nonumber\\
&\quad+n^{1/2}\mathbb{P}\Big\{\phi^*(Y,Z;\theta_0,\widehat{\eta},\widehat{F})-\phi^*(Y,Z;\theta_0,\eta_0,F_0)\Big\}\nonumber\\
&\quad+\frac{1}{2} n^{1/2}(\widehat{\vartheta}_F-\theta_0)^{\mathrm{T}}\mathbb{P}_n\phi^*_{\theta\theta}(Y,Z;\widetilde{\theta},\widehat{\eta},\widehat{F})(\widehat{\vartheta}_F-\theta_0)+o_p(1),\label{eq:thm2-full-1}
\end{align}
where $\widetilde{\theta}$ is some value between $\theta_0$ and $\widehat{\vartheta}_F$. By Condition 4, the third term on the right-hand side of (\ref{eq:thm2-full-1}) is $O(n^{1/2}r_n)$, so we have $\Vert\widehat{\vartheta}_F-\theta_{0}\Vert=O_p(n^{-1/2}+r_n)$. Likewise, we can conclude that $\Vert\widehat{\vartheta}_S-\theta_{0}\Vert=O_p(n^{-1/2}+r_n)$.

Further evaluation of the second-order residual in (\ref{eq:thm2-full-1}) yields
\begin{align}
0 & =n^{1/2}\mathbb{P}_n\phi^*(Y,Z;\theta_0,\eta_0,F_0)+n^{1/2}(\widehat{\vartheta}_F-\theta_{0})^{\mathrm{T}}\bigg\{\mathbb{P}\phi^*_\theta(Y,Z;\theta_0,\eta_0,F_0)+o_p(1)\bigg\}\nonumber\\
&\quad+ n^{1/2}\mathbb{P}\Big\{\phi^*(Y,Z;\theta_0,\widehat{\eta},\widehat{F})-\phi^*(Y,Z;\theta_0,\eta_0,F_0)\Big\}\nonumber\\
&\quad+\frac{1}{2} n^{1/2}(\widehat{\vartheta}_F-\theta_0)^{\mathrm{T}}\mathbb{P}\phi^*_{\theta\theta}(Y,Z;\theta_0,\eta_0,F_0)(\widehat{\vartheta}_F-\theta_0)\nonumber\\
&\quad+\frac{1}{2} n^{1/2}(\widehat{\vartheta}_F-\theta_0)^{\mathrm{T}}\mathbb{P}\Big\{\phi^*_{\theta\theta}(Y,Z;\widetilde{\theta},\widehat{\eta},\widehat{F})-\phi^*_{\theta\theta}(Y,Z;\widetilde{\theta},\eta_0,F_0)\Big\}(\widehat{\vartheta}_F-\theta_0)\nonumber\\
&\quad+O_p(n^{1/2}\Vert\widehat{\vartheta}_F-\theta_0\Vert^3)+o_p(1).\label{eq:thm2-full-2}
\end{align}
Note that the last three terms on the right-hand side of (\ref{eq:thm2-full-2}) are $o_p(1)$.

Likewise, for the subsample estimator, we have
\begin{align}
0 & =n^{1/2}\mathbb{P}_n\bigg\{\frac{R}{\pi}\phi^*(Y,Z;\widehat{\vartheta}_S,\widehat{\eta},\widehat{F})\bigg\}\nonumber\\
&=n^{1/2}\mathbb{P}_n\bigg\{\frac{R}{\pi}\phi^*(Y,Z;\theta_0,\eta_0,F_0)\bigg\}+n^{1/2}(\widehat{\vartheta}_S-\theta_{0})^{\mathrm{T}}\bigg\{\mathbb{P}\phi^*_\theta(Y,Z;\theta_0,\eta_0,F_0)+o_p(1)\bigg\}\nonumber\\
 &\quad+n^{1/2}\mathbb{P}\Big\{\phi^*(Y,Z;\theta_0,\widehat{\eta},\widehat{F})-\phi^*(Y,Z;\theta_0,\eta_0,F_0)\Big\}\nonumber\\
&\quad+\frac{1}{2} n^{1/2}(\widehat{\vartheta}_S-\theta_0)^{\mathrm{T}}\mathbb{P}\phi^*_{\theta\theta}(Y,Z;\theta_0,\widehat{\eta},\widehat{F})(\widehat{\vartheta}_S-\theta_0)+o_p(1).\label{eq:thm2-sub}
\end{align}
Subtracting (\ref{eq:thm2-full-2}) from (\ref{eq:thm2-sub}), we have that
\begin{align*}
n^{1/2}\mathbb{P}_n\bigg\{\bigg(\frac{R}{\pi}-1\bigg)\phi^*(Y,Z;\theta_0,\eta_0,F_0)\bigg\}+n^{1/2}(\widehat{\vartheta}_S-\widehat{\vartheta}_{F})^{\mathrm{T}}\mathbb{P}\phi^*_\theta(Y,Z;\theta_0,\eta_0,F_0)+o_p(n^{1/2}\Vert\widehat{\vartheta}_S-\widehat{\vartheta}_F\Vert+1)
\end{align*}
equals zero.
We conclude that $n^{1/2}\Vert\widehat{\vartheta}_S-\widehat{\vartheta}_F\Vert=O_p(1)$. As a result, we have
\[
n^{1/2}(\widehat{\vartheta}_S-\widehat{\vartheta}_F)=\Big\{-\mathbb{P}\phi^*_\theta(Y,Z;\theta_0,\eta_0,F_0)\Big\}^{-1}n^{1/2}\mathbb{P}_n\bigg\{\bigg(\frac{R}{\pi}-1\bigg)\phi^*(Y,Z;\theta_0,\eta_0,F_0)\bigg\}+o_p(1).
\]
The results follow from the proof of Theorem 1.
\end{proof}

\begin{proof}[Proof of Theorem 3]
It suffices to prove the results for $q=2$. The first inequality is simply a result of Theorem 1. To simplify expressions, let
\[
V_{jk}=\mathrm{E}\bigg(\frac{1-\pi}{\pi}\phi_{0j}\phi_{0k}^{\mathrm{T}}\bigg)
\]
for $j,k=0,1,2$, where $\phi_{0k}=\phi_k(Y,Z;\theta_0,\eta_0,F_0)$ for $k=1,2$, and $\phi_{00}=\psi_0$. Note that
\[
n^{1/2}(\widehat{\theta}^{\mathrm{(Seq)}}_1-\theta_0)=n^{1/2}\mathbb{P}_n\bigg\{\frac{R}{\pi}\psi_0-V_{01}V_{11}^{-1}\bigg(\frac{R}{\pi}-1\bigg)\phi_{01}\bigg\}+o_p(1),
\]
and $\Sigma^{\mathrm{(Seq)}}_{12,2}=V_{02}-V_{01}V_{11}^{-1}V_{12}$. We have
\begin{align*}
n^{1/2}(\widehat{\theta}^{\mathrm{(Seq)}}_2-\theta_0)=&\,n^{1/2}\mathbb{P}_n\bigg\{\frac{R}{\pi}\psi_0-V_{01}V_{11}^{-1}\bigg(\frac{R}{\pi}-1\bigg)\phi_{01}-(V_{02}-V_{01}V_{11}^{-1}V_{12})V_{22}^{-1}\bigg(\frac{R}{\pi}-1\bigg)\phi_{02}\bigg\}+o_p(1),
\end{align*}
We can then derive that $\Sigma^{\mathrm{(Seq)}}_2$ is equal to
\begin{align*}
 & V-2V_{01}V_{11}^{-1}V_{10}-2(V_{02}-V_{01}V_{11}^{-1}V_{12})V_{22}^{-1}V_{20}+V_{01}V_{11}^{-1}V_{10}\\
 &\quad+2(V_{02}-V_{01}V_{11}^{-1}V_{12})V_{22}^{-1}V_{21}V_{11}^{-1}V_{10}\\
 & \quad+(V_{02}-V_{01}V_{11}^{-1}V_{12})V_{22}^{-1}(V_{20}-V_{21}V_{11}^{-1}V_{10})\\
=\, & V-V_{01}V_{11}^{-1}V_{10}-(V_{02}-V_{01}V_{11}^{-1}V_{12})V_{22}^{-1}(V_{20}-V_{21}V_{11}^{-1}V_{10}),
\end{align*}
where $V=\mathrm{E}\big(\pi^{-1}\psi_0\psi_0^{\mathrm{T}}\big)$.
Since $\Sigma^{\mathrm{(Seq)}}_1=V-V_{01}V_{11}^{-1}V_{10}$, we have $\Sigma_2^{\mathrm{(Seq)}}\preceq\Sigma_1^{\mathrm{(Seq)}}$.

Now, note that $\Sigma^{\mathrm{(Joint)}}$ is equal to
\begin{align*}
 & V - \left(\begin{array}{cc}
V_{01} & V_{02}\end{array}\right)\left(\begin{array}{cc}
V_{11} & V_{12}\\
V_{21} & V_{22}
\end{array}\right)^{-1}\left(\begin{array}{c}
V_{10}\\
V_{20}
\end{array}\right)\\
=\, & V-\left(\begin{array}{cc}
V_{01} & V_{02}\end{array}\right)\left(\begin{array}{cc}
V_{11}^{-1}+V_{11}^{-1}V_{12}\widetilde{V}_{22}^{-1}V_{21}V_{11}^{-1} & -V_{11}^{-1}V_{12}\widetilde{V}_{22}^{-1}\\
-\widetilde{V}_{22}^{-1}V_{21}V_{11}^{-1} & \widetilde{V}_{22}^{-1}
\end{array}\right)\left(\begin{array}{c}
V_{10}\\
V_{20}
\end{array}\right)\\
=\, & V - V_{01}V_{11}^{-1}V_{10}-(V_{02}-V_{01}V_{11}^{-1}V_{12})\widetilde{V}_{22}^{-1}(V_{20}-V_{21}V_{11}^{-1}V_{10}),
\end{align*}
where $\widetilde{V}_{22}=V_{22}-V_{21}V_{11}^{-1}V_{12}$.
Since $V_{22}^{-1}\preceq\widetilde{V}_{22}^{-1}$, we have $\Sigma^{\mathrm{(Joint)}}\preceq\Sigma^{\mathrm{(Seq)}}_2$.
\end{proof}

\begin{proof}[Proof of Proposition 1] Since $\phi^*$, $\phi_\theta^*$, and $\phi^*_{\theta\theta}$ have uniformly bounded derivatives with respect to $\theta$, $\{g(Y,Z;\theta,F_0):\theta\in\Theta\}$ is Donsker for $g=\phi^*$, $\phi^*_\theta$, and $\phi^*_{\theta\theta}$, so the first equation of Condition 4 holds. For the second equation, consider $g=\phi^*$. We can write
\begin{align*}
\phi^*(y,z;\theta,\widehat{F})=&\,\frac{n^{-1}\sum_{j=1}^na_n^{-p}R_j\pi_j^{-1}\psi(y,X_j,z;\theta)f(y\mid X_j,z;\theta)\prod_{k=1}^pK((Z_{jk}-z_k)/a_n)}{n^{-1}\sum_{j=1}^na_n^{-p}R_j\pi_j^{-1}f(y\mid X_j,z;\theta)\prod_{k=1}^pK((Z_{jk}-z_k)/a_n)}\\\equiv&\, \frac{n^{-1}\sum_{j=1}^n A_n((y,z),(R_j,\pi_j,X_j,Z_j);\theta)}{n^{-1}\sum_{j=1}^n B_n((y,z),(R_j,\pi_j,X_j,Z_j);\theta)}.
\end{align*}
Let
\[
\phi_n^*(y,z;\theta)=\frac{\mathrm{E}\{A_n((y,z),(R,\pi,X,Z);\theta)\}}{\mathrm{E}\{B_n((y,z),(R,\pi,X,Z);\theta)\}};
\]
note that $\pi$ is a function of $(Y,Z)$. We can write
\begin{align}
\phi^*(y,z;\theta,\widehat{F}) - \phi^*(y,z;\theta,F_0)=\{\phi^*(y,z;\theta,\widehat{F}) - \phi_n^*(y,z;\theta)\} + \{\phi_n^*(y,z;\theta) - \phi^*(y,z;\theta,F_0)\}.\label{eq:phidiff}
\end{align}
To bound the second difference on the right-hand side of (\ref{eq:phidiff}), note that
\begin{align*}
\mathrm{E}\{B_n((y,z),(R,\pi,X,Z);\theta)\}=&\,\int\mathrm{E}\bigg(\frac{R}{\pi}f(y\mid X,z;\theta)\mid Z=z'\bigg)f_Z(z')\frac{1}{a_n^p}\prod_{k=1}^pK\bigg(\frac{z_k-z'_k}{a_n}\bigg)\,\mathrm{d}z'\\
=&\,\int\mathrm{E}\bigg(\frac{R}{\pi}f(y\mid X,z;\theta)\mid Z=z+a_nu\bigg)f_Z(z+a_nu)\prod_{k=1}^pK(u_k)\,\mathrm{d}u\\
=&\, \mathrm{E}(f(y\mid X,z;\theta)\mid Z=z)f_Z(z)+O(a_n^2).
\end{align*}
Likewise,
\[
\mathrm{E}\{A_n((y,z),(R,\pi,X,Z);\theta)\}=
\mathrm{E}(\psi(y,X,z;\theta)f(y\mid X,z;\theta)\mid Z=z)f_Z(z)+O(a_n^2).
\]
Therefore, the second difference on the right-hand side of (\ref{eq:phidiff}) is $O(a_n^2)=o(1)$. For the first difference,
\begin{align*}
&\,\phi^*(y,z;\theta,\widehat{F}) - \phi^*_n(y,z;\theta)\\
=&\, \frac{1}{n} \sum_{j=1}^n \Bigg\{ 
    \frac{A_n((y,z),(R_j,\pi_j,X_j,Z_j);\theta) - \mathrm{E}A_n((y,z),(R,\pi,X,Z);\theta)}
         {\mathrm{E}B_n((y,z),(R,\pi,X,Z);\theta)} - \frac{
    \mathrm{E} A_n((y,z),(R,\pi,X,Z);\theta)
}{
    \left\{\mathrm{E} B_n((y,z),(R,\pi,X,Z);\theta)\right\}^2
}
\\
&\qquad \times\left[ B_n((y,z),(R_j,\pi_j,X_j,Z_j);\theta) - \mathrm{E}B_n((y,z),(R,\pi,X,Z);\theta) \right]
\Bigg\}+r_{n1} \\
=&\,\frac{1}{n} \sum_{j=1}^n \Bigg\{ 
    \frac{A_n((y,z),(R_j,\pi_j,X_j,Z_j);\theta) - \mathrm{E}A_n((y,z),(R,\pi,X,Z);\theta)}
         {\int f(y\mid x,z;\theta)\,\mathrm{d}F(x\mid z)f_Z(z)} \\
         &\qquad - \frac{
    \int \psi(y,x,z;\theta)f(y\mid x,z;\theta)\,\mathrm{d}F(x\mid z)
}{
    f_Z(z)\left[\int f(y\mid x,z;\theta)\,\mathrm{d}F(x\mid z)\right]^2
}\\
&\qquad\times
\left[ B_n((y,z),(R_j,\pi_j,X_j,Z_j);\theta) - \mathrm{E} B_n((y,z),(R,\pi,X,Z);\theta) \right]
\Bigg\}+r_{n1}+r_{n2}\\
\equiv&\,\frac{1}{n}\sum_{j=1}^ng_n((y,z),(R_j,\pi_j,X_j,Z_j);\theta)+r_{n1}+r_{n2},
\end{align*}
where $r_{n1}$ involves second-order terms of the differences $\{A_n((y,\allowbreak z),\allowbreak (R_j,\allowbreak \pi_j,\allowbreak X_j,\allowbreak Z_j);\theta) -\allowbreak \mathrm{E}A_n((y,\allowbreak z),\allowbreak (R,\allowbreak \pi,\allowbreak X,\allowbreak Z);\theta)\}$ and $\{B_n((y,\allowbreak z),\allowbreak (R_j,\allowbreak \pi_j,\allowbreak X_j,\allowbreak Z_j);\theta) -\allowbreak \mathrm{E}B_n((y,\allowbreak z),\allowbreak (R,\allowbreak \pi,\allowbreak X,\allowbreak Z);\theta)\}$, and $r_{n2}$ involves these difference terms times kernel approximation errors of order $O(a_n^2)$.

Combining the above results, we have
\begin{align*}
&\,n^{1/2}(\mathbb{P}_n-\mathbb{P})\Big\{\phi^*(Y,Z;\theta,\widehat{F})-\phi^*_n(Y,Z;\theta)\Big\}\\
=&\,n^{1/2}\bigg\{\frac{1}{n^2}\sum_{i=1}^n\sum_{j=1}^ng_n((Y_i,Z_i),(R_j,\pi_j,X_j,Z_j);\theta)-\frac{1}{n}\sum_{i=1}^n\mathrm{E}_{(R,\pi,X,Z)}g_n((Y_i,Z_i),(R,\pi,X,Z);\theta)\\
&\quad-\frac{1}{n}\sum_{j=1}^n\mathrm{E}_{(Y,Z)}g_n((Y,Z),(R_j,\pi_j,X_j,Z_j);\theta)+\mathrm{E}g_n((Y,Z),(R',\pi',X',Z');\theta)\bigg\}+r_{n3}\\
\equiv&\,n^{1/2}\bigg\{\frac{1}{n^2}\sum_{i=1}^n\sum_{j=1}^ng_{12,n}((Y_i,Z_i),(R_j,\pi_j,X_j,Z_j);\theta)\bigg\}+r_{n3},
\end{align*}
where $r_{n3}$ is a residual term that converges to zero faster than the first term on the right-hand side above, and $(Y,R,\pi,X,Z)$ and $(R',\pi',X',Z')$ are copies of the variables that are independent of the observed data and of each other. Let
\[
\mathcal{G}_n=\Big\{g_{12,n}(\cdot,\cdot;\theta):\theta\in\Theta\Big\}.
\]
Under Condition 4B, we can find an envelope function of $\mathcal{G}_n$ of the order $a_n^{-p}$, denoted by $G_n$. Let $\mathcal{F}_n$ denote the class of functions
\[
\mathcal{F}_n = \left\{ (y, z), (y', x', z') \mapsto \psi(y, x', z; \theta)^\delta \frac{1}{\pi(y',z')}f(y \mid x', z; \theta) \prod_{k=1}^pK\left(\frac{z'_k - z_k}{a_n}\right) : \theta \in \Theta,\delta\in\{0,1\} \right\}.
\]
Since $\psi$ and $f$ are smooth in $\theta$ and $\Theta$ is compact, $\mathcal{F}_n$ is a parametric class and its covering number satisfies
\[
\log N(\epsilon, \mathcal{F}_n, L_2(\mathbb{Q}^2)) \leq C \{1+\log(1/\epsilon)\}
\]
for $\epsilon\in(0,1]$, where $\mathbb{Q}$ can be any probability measure, and $C$ is a constant. Note that $\mathcal{G}_n$ consists of linear combinations of functions in $\mathcal{F}_n$ scaled by $a_n^{-p}$, so the covering number rescales as
\[
N(\epsilon, a_n^{-p} \mathcal{F}_n, L_2(\mathbb{Q}^2)) = N(a_n^p \epsilon, \mathcal{F}_n, L_2(\mathbb{Q}^2)).
\]
Therefore,
\[
\log N(\epsilon, \mathcal{G}_n, L_2(\mathbb{Q}^2)) \leq C\left\{1+ \log\left(\frac{1}{a_n^p \epsilon}\right)\right\}.
\]

Now, we wish to bound
\begin{align}
\mathrm{E}\bigg\Vert\sum_{i=1}^n\sum_{j=1}^ng_{12,n}((Y_i,Z_i),(R_j,\pi_j,X_j,Z_j);\theta)\bigg\Vert_{\mathcal{G}_n}\label{eq:uprocess}
\end{align}
using Theorem 6 of \cite{nolan1987u}. Let $\mathbb{T}_n$ be a discrete measure on $(\mathcal{Y},\mathcal{Z})\times(\{0,1\},\pi(\mathcal{Y},\mathcal{Z}),\mathcal{X},\mathcal{Z})$ defined according to $T_n$ in \citet[pp.782]{nolan1987u}, which assigns a point mass of 1 at $4n(n-1)$ pairs of $(Y_i,Z_i)_{i=1,\ldots,n}\times(R_j,\pi_j,X_j,Z_j)_{j=1,\ldots,n}$.
Define a distance measure $d_{\mathbb{T}_n,2,G_n}$ such that for $g_1,g_2\in\mathcal{G}_n$,
\[
d_{\mathbb{T}_n,2,G_n}(g_1,g_2)=\bigg\{\frac{\mathbb{T}_n|g_1-g_2|^2}{\mathbb{T}_nG_n^2}\bigg\}^{1/2}.
\]
Since the functions in $\mathcal{G}_n$ and their envelope $G_n$ share the common scale $a_n^{-p}$, this scale cancels in the envelope-normalized metric $d_{\mathbb{T}_n,2,G_n}$. Therefore,
\[
\log N(\epsilon,\mathcal{G}_n,d_{\mathbb{T}_n,2,G_n})\lesssim1+\log\bigg(\frac{1}{\epsilon}\bigg),
\]
so
\[
J_n(s)\equiv\int_0^s\log N(\epsilon,\mathcal{G}_n,d_{\mathbb{T}_n,2,G_n})\,\mathrm{d}\epsilon\lesssim s\log(e/s)
\]
for $s\in(0,1]$. Also, we have $\tau_n\equiv(\mathbb{T}_nG_n^2)^{1/2}\asymp na_n^{-p}$ and
\[
\theta_n\equiv\frac{1}{4}\sup_{g\in\mathcal{G}_n}(\mathbb{T}_ng^2)^{1/2}=O(na_n^{-p}).
\]
Clearly, $\theta_n/\tau_n\le1/4$, so $J_n(\theta_n/\tau_n)$ is bounded. By Theorem 6 of \cite{nolan1987u},
\[
\mathrm{E}\bigg\Vert\sum_{i=1}^n\sum_{j=1}^ng_{12,n}((Y_i,Z_i),(R_j,\pi_j,X_j,Z_j);\theta)\bigg\Vert_{\mathcal{G}_n}\lesssim\mathrm{E}\Big\{\theta_n+\tau_nJ_n(\theta_n/\tau_n)\Big\}\lesssim na_n^{-p}.
\]
Therefore,
\[
\sup_{\theta\in\Theta}\bigg|n^{1/2}(\mathbb{P}_n-\mathbb{P})\Big\{\phi^*(Y,Z;\theta,\widehat{F})-\phi^*(Y,Z;\theta,F_0)\Big\}\bigg|=O_p(n^{-1/2}a_n^{-p})+O(a_n^2),
\]
which converges to 0 in probability under Condition 4A. The second equation of Condition 4 with $g=\phi^*_\theta$ or $\phi^*_{\theta\theta}$ is proved similarly.

For the third equation of Condition 4, by standard arguments for kernel methods based on the decomposition in (\ref{eq:phidiff}), we can conclude that
\[
\mathbb{P}\Big\{\phi^*(Y,Z;\theta,\widehat{F})-\phi^*(Y,Z;\theta,F_0)\Big\}=O_p(n^{-1/2}+a_n^2)=O_p(r_n),
\]
where $r_n$ satisfies $n^{1/2}r_n^3\rightarrow0$.
\end{proof}

\clearpage
\bibliographystyle{apalike}
\bibliography{ref}
\end{document}